%% file: main.tex
\documentclass{vldb}

\input{commands}

\usepackage{diagbox}
\usepackage{makecell}

\vldbTitle{Reusing Machine Learning Classifiers for Entity Resolution by Transfer Learning}
\vldbAuthors{}
\vldbDOI{https://doi.org/TBD}
\vldbVolume{12}
\vldbNumber{xxx}
\vldbYear{2019}

\makeatletter
\def\@copyrightspace{\relax}
\makeatother

\begin{document}

\title{Reusing Machine Learning Classifiers \\ for Entity Resolution by Transfer Learning}
\title{Reuse and Adaptation for Entity Resolution \\through Transfer Learning}

\author{
	Saravanan Thirumuruganathan$^\ddag$\hspace{3ex}
	Shameem A Puthiya Parambath$^\ddag$ \\
	Mourad Ouzzani$^\ddag$\hspace{3ex}
	Nan Tang$^\ddag$\hspace{3ex}
	Shafiq Joty$^\dag$\\
	\affaddr{$^\ddag$Qatar Computing Research Institute, HBKU, Qatar \hspace{2ex} $^\dag$Nanyang Technological University, Singapore}
	\\
	\{sthirumuruganathan, spparambath, mouzzani, ntang\}@hbku.edu.qa \hspace{3ex} srjoty@ntu.edu.sg
}

\maketitle

\begin{abstract}
Entity resolution (ER) is one of the fundamental problems in data integration, where machine learning (ML) based classifiers often provide the state-of-the-art results. Considerable human effort goes into feature engineering and training data creation. In this paper, we investigate a new problem: 
{\it Given a dataset $D_T$ for ER with limited or no training data, is it possible to train a good ML classifier on $D_T$ by reusing and adapting the training data of dataset $D_S$ from same or related domain?}
Our major contributions include 
(1) a distributed representation based approach to encode each tuple from diverse datasets into a standard feature space; 
(2) identification of common scenarios where the reuse of training data can be beneficial; and (3) five algorithms for handling each of the aforementioned scenarios. 
We have performed comprehensive experiments on {12 datasets from 5} different domains (publications, movies, songs, restaurants, and books). Our experiments show that our algorithms provide significant benefits such as providing superior performance {for a fixed training data size}.
\end{abstract}

\input{sec-introduction}
\input{sec-preliminaries}
\input{sec-featurenormalization}

\input{sec-algorithms}

\input{sec-putting-together}
\input{sec-extensions}

\input{sec-expt}
\input{sec-relWork}

\input{sec-conclusion}

\bibliographystyle{abbrv}
\bibliography{DA}

\end{document}

%% file: commands.tex
\newcommand{\eat}[1]{}

\usepackage{latexsym}
\usepackage{amsfonts}
\usepackage{amsmath}
\usepackage{amssymb}
\usepackage{xcolor}
\usepackage{colortbl}
\usepackage{epsfig}
\usepackage{xspace}
\usepackage{graphicx}
\usepackage{subfigure}
\usepackage{paralist}
\usepackage{enumerate}
\usepackage[color,matrix,arrow,all]{xy}
\usepackage{comment}
\usepackage{booktabs}
\usepackage{balance}
\usepackage{stmaryrd}
\usepackage{pifont}
\usepackage{hhline}
\usepackage{listings}
\usepackage{array}
\usepackage{float}
\usepackage[flushleft]{threeparttable}

\usepackage{wrapfig}

\newcolumntype{L}[1]{>{\raggedright\let\newline\\\arraybackslash\hspace{0pt}}m{#1}}
\newcolumntype{C}[1]{>{\centering\let\newline\\\arraybackslash\hspace{0pt}}m{#1}}
\newcolumntype{R}[1]{>{\raggedleft\let\newline\\\arraybackslash\hspace{0pt}}m{#1}}

\usepackage{epsfig}
\usepackage{multirow}
\usepackage{url}

\usepackage[export]{adjustbox}
\usepackage[noend]{algorithmic}
\usepackage{algorithm}

\newcommand{\add}[1]{\textcolor{blue}{{#1}}}

\newcommand{\at}[1]{\protect\ensuremath{\mathsf{#1}}\xspace}

\newcommand{\sstab}{\rule{0pt}{8pt}\\[-2.2ex]}

\newcommand{\bi}{\begin{itemize}}
\newcommand{\ei}{\end{itemize}}

\newcommand{\be}{\begin{enumerate}}
\newcommand{\ee}{\end{enumerate}}
\newcommand{\beqn}{\begin{eqnarray*}}
\newcommand{\eeqn}{\end{eqnarray*}}

\newcommand{\stitle}[1]{\vspace{1ex}\noindent{\bf #1}}

\newcommand{\etitle}[1]{\vspace{0.8ex}\noindent{\underline{\em #1}}}

\newcommand{\ie}{{\em i.e.,}\xspace}
\newcommand{\eg}{{\em e.g.,}\xspace}

\newcommand{\aka}{\emph{a.k.a.}\xspace}

\newcounter{ccc}

\newcommand{\DRs}{{DRs}\xspace}
\newcommand{\DR}{{DR}\xspace}

\DeclareMathAlphabet{\pazocal}{OMS}{zplm}{m}{n}

\newcommand{\eop}{\hspace*{\fill}\mbox{$\Box$}\vspace{1ex}}     

\renewcommand{\ni}{\noindent}

\newcommand{\nthesection}{\arabic{section}}
\newcounter{alg}[section]
\renewcommand{\thealg}{\nthesection.\arabic{alg}}

\newcounter{arule}
\renewcommand{\thearule}{\arabic{arule}}

\newcounter{claim}
\renewcommand{\theclaim}{\arabic{claim}}

\makeatletter
    \newcommand\figcaption{\def\@captype{figure}\caption}
    \newcommand\tabcaption{\def\@captype{table}\caption}
\makeatother

\definecolor{shadecolor}{RGB}{200,200,200}
\newcommand{\mybox}[1]{\noindent\colorbox{shadecolor}
	{\parbox{\dimexpr\columnwidth-2\fboxsep\relax}{#1}}}

\definecolor{shadecolor1}{RGB}{230,230,230}

\definecolor{shadecolor1}{RGB}{255, 114, 118}


%% file: sec-introduction.tex
\section{Introduction}
\label{sec:introduction}

Entity resolution (ER) 
-- which identifies pairs of duplicate entities --
is a fundamental  problem in data integration. 
Several studies~\cite{konda2016magellan, anhaisigmod2018, deeper} show that machine learning (ML)-based methods
often provide state-of-the-art results for ER.

The seedy underbelly of such superior performance is the considerable human effort that underlies these ML methods.
Critical prerequisites to obtain the aforementioned state-of-the-art results often include 
the availability of enough labeled data in the form of matching and non-matching tuple pairs, 
good feature engineering, and fine-tuning of the models.
Each of these steps requires considerable human effort, which is obviously expensive.
Furthermore, many of the sophisticated ML models require large training data to achieve good results~\cite{erhard2010}. 
In fact, it has been reported~\cite{dong2018data} that 
achieving F-measures of $\sim$99\% with random forests can require 
up to 1.5M labels even for relatively clean datasets.

Large organizations often have 100s of datasets to be integrated and deduplicated~\cite{stonebraker2018data}.
Furthermore, organizations continuously produce or acquire new datasets,
and each time they have to identify duplicates before feeding these datasets to downstream processes.
Unfortunately, they need to spend a lot of human effort with each new dataset if the previous steps are all repeated from scratch.

%

\stitle{Reuse and Adaptation for ER.}
In this paper, we investigate the following problem for ER: 

\mybox{Given a target dataset $D_T$ to be deduplicated with limited or no training data, 
is it possible to train a good ML classifier for $D_T$
by reusing and adapting the training data from a related dataset $D_S$?}
%

An affirmative answer to the above problem has the potential to dramatically reduce the human effort needed.
We would like to note that our work is orthogonal and can be used in conjunction 
with other approaches to reduce 
labeling effort such as active learning~\cite{SarawagiB02} and weak supervision~\cite{ratner2016data}.



\begin{figure}[t!]
     \vspace*{-2ex}
     \includegraphics[width=\columnwidth]{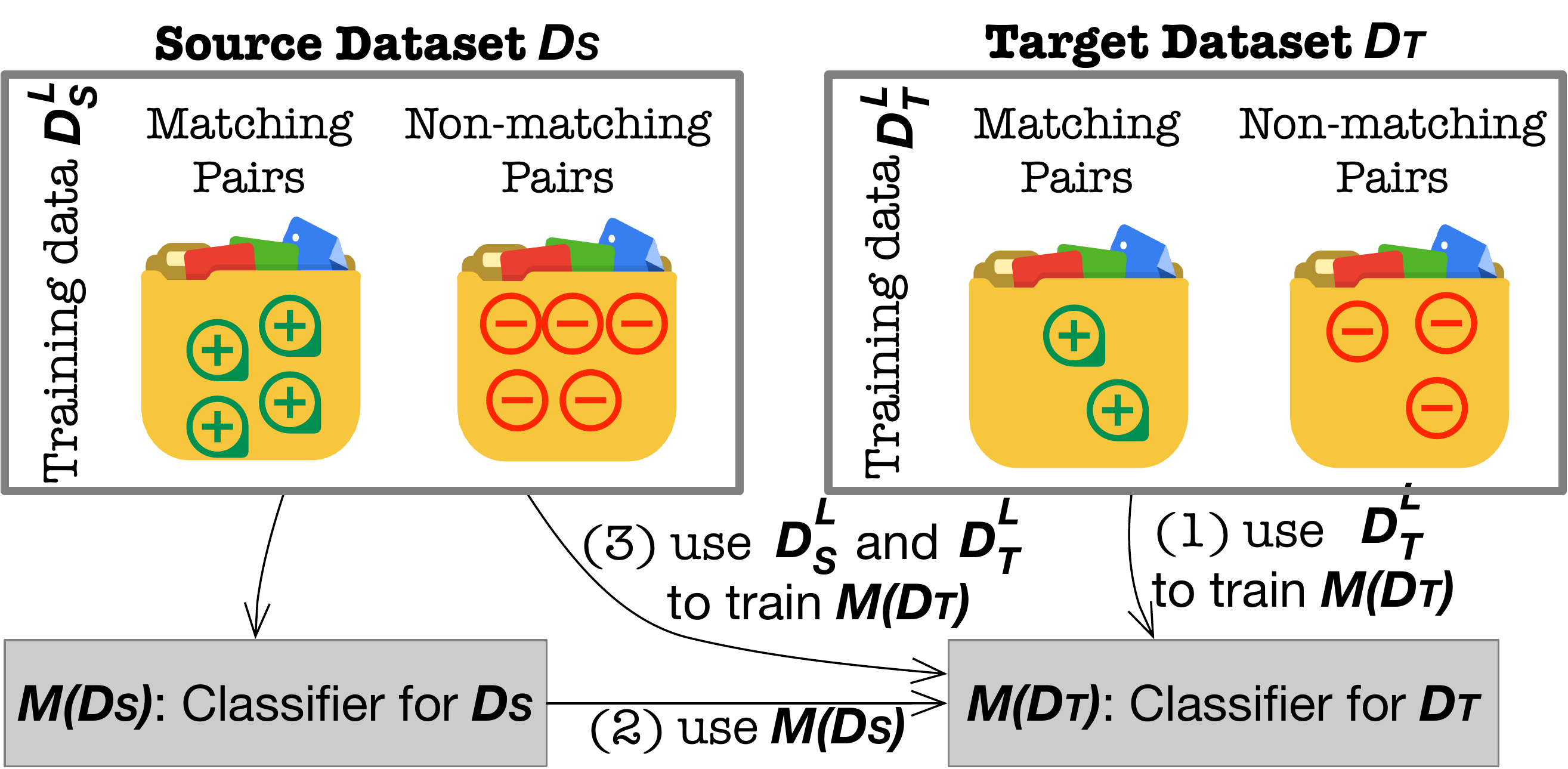}
     \vspace*{-5ex}
     \caption{Illustration of Designing ER Classifiers}
     \label{fig:ertl}
     \vspace*{-4ex}
 \end{figure}

\stitle{Prior Art and Their Limitations.}
Consider source $D_S$ and target $D_T$ datasets, with labeled training data  $D_S^L$ and $D_T^L$, respectively, where ``$+/-$'' means a matching/non-matching tuple pair, as shown in Figure~\ref{fig:ertl}.
Consider three classical methods for designing an ML classifier, $M(D_T)$, for $D_T$.
(1) {\em No Transfer (NoT)} only uses the training data $D_T^L$ from the target to train $M(D_T)$, while ignoring the training data, $D_S^L$, from the source.
(2) {\em Naive Transfer (NvT).}
The other end of the spectrum is to blindly apply the classifier built on $D_S$ to $D_T$.
%
(3) {\em Transfer learning (TL).}
A smarter approach is to use the training data not only from the target ($D_T^L$), but also  from the source ($D_S^L$), 
which is the chief goal of 
transfer learning (TL)~\cite{pan2010survey, sugiyama2017dataset}.

Obviously, (1) NoT will not work when $D_T$ has no training dataset (\ie $D_T^L = \emptyset$), or $D_T^L$ is very small thus producing a classifier which will be biased and likely overfit, especially for high capacity ML classifiers such as deep learning, random forests, and SVMs.
(2) NvT works well only if the two datasets are very similar in terms of {\em both} features {\em and} data distribution, which are 
often violated in practice, for example when we have different schemas and distributions of duplicates between the source and the target.
Although intuitive, as we shall elaborate in Section~\ref{sec:algorithms} and in the experiments, directly applying prior TL methods (3) is doomed to fail 
   since for ER purposes, datasets $D_S$ and $D_T$ might have a number of differences that make knowledge transfer challenging. 
    These can be major such as different schemata, or subtle such as different vocabulary, 
    different distributions of duplicates, 
    different duplicate to non-duplicate ratios,
    and divergence in the most discriminative features.


%
	

\stitle{Research Challenges.}
Let $X$ be the similarity vector between two tuples and $y$ be the label (1 for duplicate or 0 for not).
The Fellegi-Sunter, a central model for solving the ER problem~\cite{theoryrecordlinkage}, can be considered as a probabilistic decision rule 
that declares a given $X$ as a duplicate if the likelihood $P(y|X)$ exceeds a given threshold.
Any ML classifier can be used to learn this probability from the data. 
Given this setting, the fundamental challenges of TL for ER are:


\emph{(I) Prior Probability Shift} 
	occurs when the probability distribution of the labels vary between $D_S$ and $D_T$.
	For example, $D_S$ could be a relatively clean dataset with very few duplicates
	while $D_T$ could be a noisy dataset where a much larger fraction of tuple pairs are duplicates
	resulting in very different prior probabilities $P_S(y)$ and $P_T(y)$.

\emph{(II) Covariate Shift}
	occurs when $D_S$ and $D_T$ have very different distributions of similarity vectors $P_S(X)$ and $P_T(X)$.
	For instance, if $D_S$ is mostly clean, then most $X$ from $P_S(X)$ might have low similarity values.

\emph{(III) Class Conditional Probability Distribution Shift}
	occurs when the quantities $P_S(y|X)$ and $P_T(y|X)$ differ,
	\ie given the same similarity vector $X$, the likelihood that it is a duplicate might vary between $D_S$ and $D_T$.

\emph{(IV) Sample Selection Bias}
	occurs when the process by which datasets $D_S$ and $D_T$ are obtained is very different.
	One common culprit is blocking that can be considered as a non-random process 
	that selects a subset of all possible tuple pairs (\ie candidate set) using a \emph{label dependent} process
	where (potential) duplicates are preferentially chosen over (potential) non-duplicates.
	If $D_S$ and $D_T$ use different blocking processes (such as varying aggressiveness for pruning), 
	 the generated candidate sets could have selection bias.
	Covariate shift is a special case of sample selection bias.
	
\emph{(V) Imbalanced Data.}
	ER is usually an imbalanced classification problem as non-duplicates out-number duplicates by  orders of magnitude.
	This issue is compounded by the fact that the sizes of $D_S$ and $D_T$ 
	(or their corresponding labeled subsets $D_S^L, D_T^L$) could be imbalanced as well.
	For example, $D_S$ might have millions of tuple pairs while $D_T$ could have only thousands.
	If we blindly apply TL, any signal from $D_T$ will be overwhelmed by $D_S$.

\stitle{Contributions.}


\etitle{(1) Identification of Common ER Reuse Scenarios.}
    We identify a set of real-world scenarios that might benefit from the reuse of ER classifiers.
    Each of these scenarios constitutes an equivalence class requiring a specific reuse technique.
	(Section~\ref{sec:background})

\etitle{(2) Feature Space Standardization.}
    A fundamental roadblock in applying TL for ER is the heterogeneity of the schemata and thereby features for the classifiers.
    We propose an effective mechanism such that tuple pairs from each dataset are encoded into the same feature space, which facilitates the reuse of classifiers.
    Our method is based on \emph{distributed representations} (\DRs), a fundamental concept in deep learning. 
    This has a number of advantages such as allowing semantic similarity, making our approach as close to off-the-shelf methods as possible. 
    (Section~\ref{sec:featureSpaceStdization})



\etitle{(3) Reuse Algorithms Spectrum.}
    We propose \textit{five} 
    algorithms that can handle most of the common labeling data scenarios in applying TL to ER and the aforementioned challenges.
    These include no transfer, naive transfer, instance weighting, feature augmentation with/without labeled data. 
    The algorithms are classifier-agnostic and can be readily adapted to 
    all of the popular ER classifiers such as random forests, logistic regression, and SVMs.
	(Section~\ref{sec:algorithms})

\etitle{(4) Practical Issues.}
    We describe a number of practical issues that one might encounter when trying to reuse ML classifiers for ER
    and provide some empirical rule-of-thumb advices that are backed by extensive experiments.
    Some of the important issues that we study include: 
    (a)~how to reuse when there are multiple possible sources,
    (b)~is it possible to reuse a dataset from an unrelated domain,
    (c)~when and when not to reuse, and
    (d)~how to select an appropriate source dataset to reuse from a set of candidates.
	(Section~\ref{sec:practicalIssues})

\etitle{(5) Experiments.}
    We conduct comprehensive experiments on {12 datasets from 5 domains} (publications, movies, songs, restaurants, and books).
    The results show that TL can provide significant benefits 
    such as providing superior performance for a fixed training dataset size.
    (Section~\ref{sec:expt})

%

%% file: sec-preliminaries.tex
\section{Background and Scenarios}
\label{sec:background}

In this section, we first introduce key concepts related to ER (Section~\ref{subsec:er}) and
then categorize the different data scenarios for applying TL to ER (Section~\ref{subsec:datascenarios}).

\subsection{Entity Resolution}
\label{subsec:er}

Let $T$ and $T'$ be two relations with aligned schema $\{A_1, A_2, \ldots, A_m\}$. 
We denote by $t[A_j]$ the value of attribute $A_j$ on tuple $t$.
The problem of entity resolution (ER) is, given all distinct tuple pairs $(t, t') \in T \times T'$, 
to determine which pairs of tuples refer to the same real-world entity.
A pair of tuples is said to match (resp. mismatch) when they refer to the same (resp. different) real-world entity.

\stitle{Blocking.} 
For efficiency reasons, ER solutions typically first run {\em blocking} methods, which generate a candidate set $C \subseteq T \times T'$ that includes tuple pairs that are likely to match.


\stitle{Training Data.}
Most, if not all, ER solutions need training data, which is formalized as follows.
A labeled training dataset is $L \subseteq T \times T' \times \{0, 1\}$,
where the triplet $(t, t', 1)$ (resp. $(t, t', 0)$) denotes that tuples $t$ and $t'$ are (resp. are not) duplicates.


\stitle{Fellegi-Sunter Model}~\cite{theoryrecordlinkage}
is a formal framework for probabilistic ER and most prior ML works are simple variants of this approach. 
Basically, they train an ML classifier $M$ on the labeled dataset $L$ such that $M$ can accurately distinguish tuple pairs in candidate set $C$ to be either match or non-match, by using vectors of similarity scores between aligned attributes as features.


\stitle{Monotonicity of Precision}
was introduced in~\cite{arasu2010active}; it states that if the similarity score of $(t_1, t_2)$ is higher than $(t_1', t_2')$, then $(t_1, t_2)$ is more likely to be a duplicate.
Although the monotonicity assumption could be violated for particular tuple pairs, it usually holds in an aggregate sense.


\subsection{ER Scenarios for TL}
\label{subsec:datascenarios}


\stitle{Transfer Learning (TL)} seeks to utilize the knowledge 
from one or more related \emph{sources}, usually in the form of  features, classifier, and training data,  in order to design an ML classifier for a \emph{target} dataset with limited or no training data
(please refer to survey~\cite{pan2010survey} and Section~\ref{sec:relWork} for more details.). 


%
%
%

In this work, we primarily seek to transfer both features (by encoding tuples in a standard feature space) and training data, based on (a)~encoding tuple pairs as a similarity vector as per the Fellegi-Sunter model, and 
(b)~assuming the monotonicity of precision.
Directly transferring ML classifiers is often much trickier and require non-trivial changes.
Our approach, that does not transfer ML classifiers, has a number of advantages such as requiring minimal changes to existing  ER pipelines 
and providing flexibility to use different classifiers between the source and target datasets.

%

\stitle{ER Scenarios for TL.}
Let the target dataset be $D_T = (T, T')$.
Let the source dataset be $D_S = (T_S, T_S')$.
We will postpone the discussion for $n$  sources $\{D_S^1=(T_1, T_1'), \ldots, D_S^n=(T_n, T_n')\}$ to Section~\ref{sec:practicalIssues}.



We categorize  ER scenarios based on the amount of training data available for building an ML classifier.
We use the terms ``\emph{limited}'' and ``\emph{adequate}'' to loosely denote that the 
amount of training data required for building an ML classifier is not sufficient and sufficient, respectively.
Of course, this can vary for each dataset and ML classifier, and is often determined by a domain expert.
For example, deep learning based methods may require much larger training data than simpler models such as decision trees.

We consider four major scenarios for {\bf (Source, Target)} that commonly occur in practice:  
%
(1) {(Adequate, Nothing)}, 
(2) {(Adequate, Limited)}, 
(3) {(Limited, Limited)}, and
(4) {(Adequate, Adequate)}.

\begin{figure}[t!]
     \vspace*{-2ex}
     \centering
     \includegraphics[width=0.8\columnwidth]{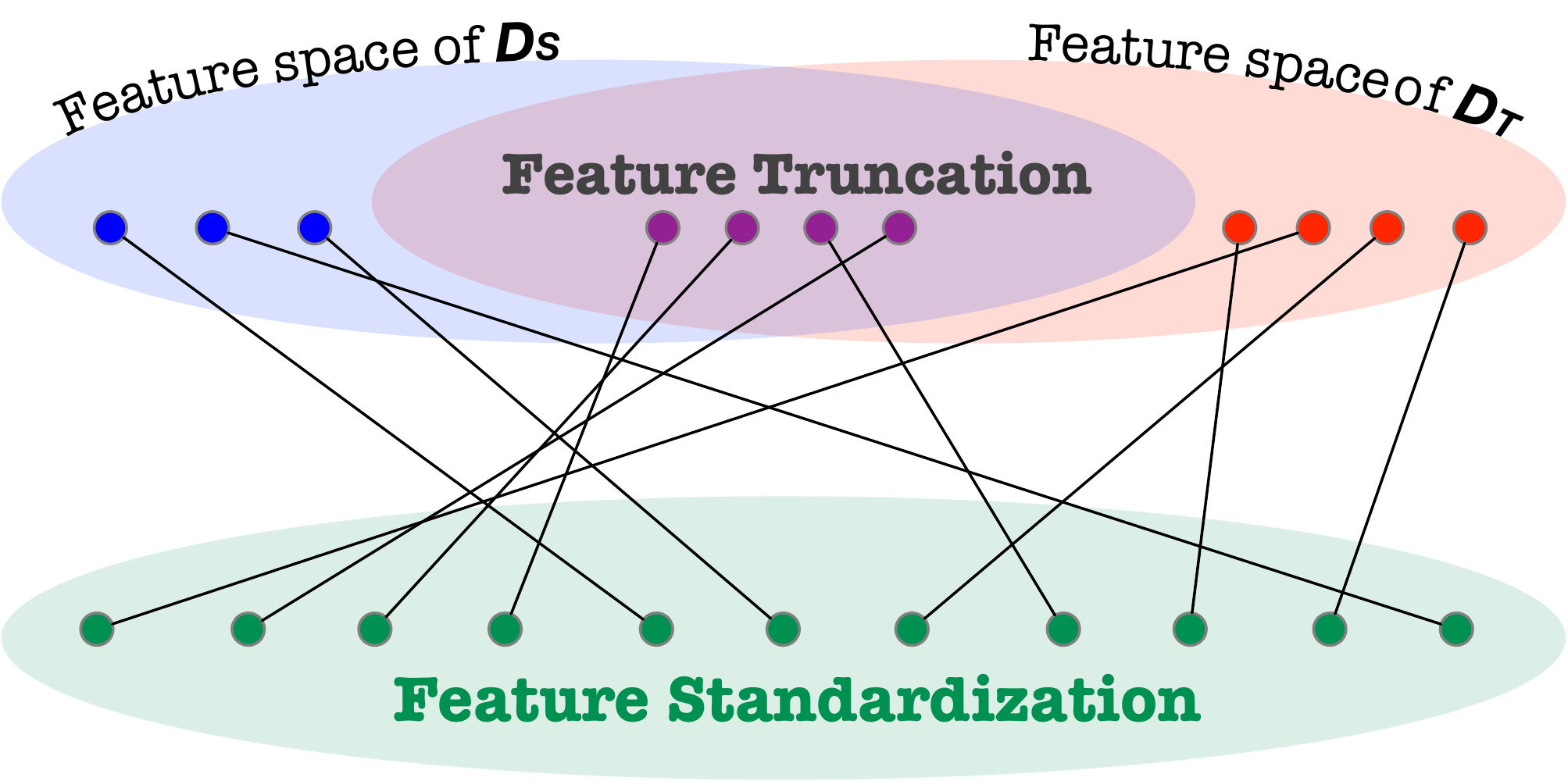}
     \vspace*{-3ex}
     \caption{Feature Truncation and Standardization}
     \label{fig:ftfs}
     \vspace*{-3.5ex}
 \end{figure}

Scenario (Adequate, Nothing) is hard as one has to build a classifier using the training data from $D_S$ 
that yet performs satisfactorily on $D_T$. 
For the scenario (Adequate, Limited), we wish to ``augment'' the limited training data from in $D_T$ with the training data from $D_S$.
This augmentation must be done in a careful manner so that the signals from the source datasets do not swamp the ones from $D_T$. 
(Limited, Limited) is a tricky scenario where we do not have adequate labeled data for both source and target datasets.
Instead of throwing up our hands in the air, we can try to ``pool'' these limited resources so that we can get an effective classifier that can perform well on each of these datasets.
One can apply a traditional ML classifier and achieve satisfactory results for the final scenario (Adequate, Adequate).
However, it is still possible to improve the performance of a classifier 
by transferring some additional relevant knowledge and make it more robust.
%

%
%
%

%% file: sec-featurenormalization.tex
\section{Feature Space Standardization}
\label{sec:featureSpaceStdization}

In this section, we first describe the rationale for encoding all the datasets into a standard feature space (Section~\ref{subsec:featureStdRationale}).  
We then propose such a feature space based on distributed representations -- a fundamental concept from deep learning. 
We also propose an algorithm to encode each tuple into this standard feature space and 
describe the various advantages conferred by such an approach (Section~\ref{subsec:drForWordsTuples}).

\subsection{Rationale for Feature Standardization}
\label{subsec:featureStdRationale}

\stitle{Feature Space Truncation vs Standardization.}
Given ``source'' $D_S$ and ``target'' $D_T$,
it is not necessary -- and even unlikely -- that they share a common schemata.
Consider an example where $D_T$ and $D_S$ have $10$ and $15$ attributes, respectively, with $7$ attributes in common.
We cannot directly use the classifier for $D_S$ to $D_T$ as they will have totally different feature spaces.
A naive approach would be to (re)build an ML classifier for $D_S$ using only the $7$ common attributes and apply it on $D_T$.
We dub this simplistic and unappealing approach as \emph{feature space truncation}.

We instead advocate for a \emph{feature space standardization} approach 
where tuples from each relation in $D_S$ and $D_T$ 
are all encoded into a standard feature space of dimension $d$. 
Then we can build an ML classifier based on the Fellegi-Sunter model, 
where the input vector measures the similarity between two tuples in this standardized (or shared) feature space.
Figure~\ref{fig:ftfs} illustrates the difference between feature truncation and feature standardization.


\stitle{Feature Space Standardization in Other Domains.}
Such standardization has been successfully used in a number of domains. 
Consider an object recognition task in the  computer vision domain on an input image with $1024 \times 1024$ pixels. 
\emph{Any} image -- whether it is that of a bird, an animal, a mountain or some random object -- can be encoded as an image with $1024 \times 1024$ pixels.
In other words, all images are implicitly encoded into the same feature space.
Hence, an ML model for classifying the given image as cats or dogs (possibly trained over images photographed indoors)
can also be used to classify cats vs. dogs from another domain (images of cats and dogs taken outdoors). 
A similar phenomenon happens in information retrieval (IR) also.
Given a fixed vocabulary -- say from the English dictionary -- it is possible to encode \emph{every} English document
as term-frequency vectors under the bag of words semantics.
An ML model trained for categorizing an email as spam vs. non-spam in one domain can be applied over another domain
as long as the inputs can be encoded in this feature space. 
One might be tempted to use the IR approach for feature space standardization in ER.
Specifically, construct a dictionary of all unique words from $D_S$ and $D_T$ and
then encode each tuple  either as a term-frequency or TF-IDF vector under the bag-of-words semantics.
However, this approach has a number of disadvantages.
First, this has to be repeated for different $D_S$ and $D_T$.
Second, this simplistic encoding does not allow the use of sophisticated similarity measures such as semantic measures.
As we shall see, our proposed approach dramatically improves on this naive way.

\subsection{Distributed Representation for Tuples}
\label{subsec:drForWordsTuples}

We propose a standard feature space by using a fundamental concept from deep learning,
namely distributed representations (\DRs) of words.
We first describe these \DRs (see~\cite{Goodfellow-et-al-2016} for more details).
We then present how to compose a \DR for a tuple from its component words.
Note that the rest of the paper is oblivious to the specific feature space and 
readily admit other approaches that could encode tuples from various datasets as a vector in a fixed feature space.

\begin{figure}[t!]
     \vspace*{-2ex}
     \centering
     \includegraphics[width=0.7\columnwidth]{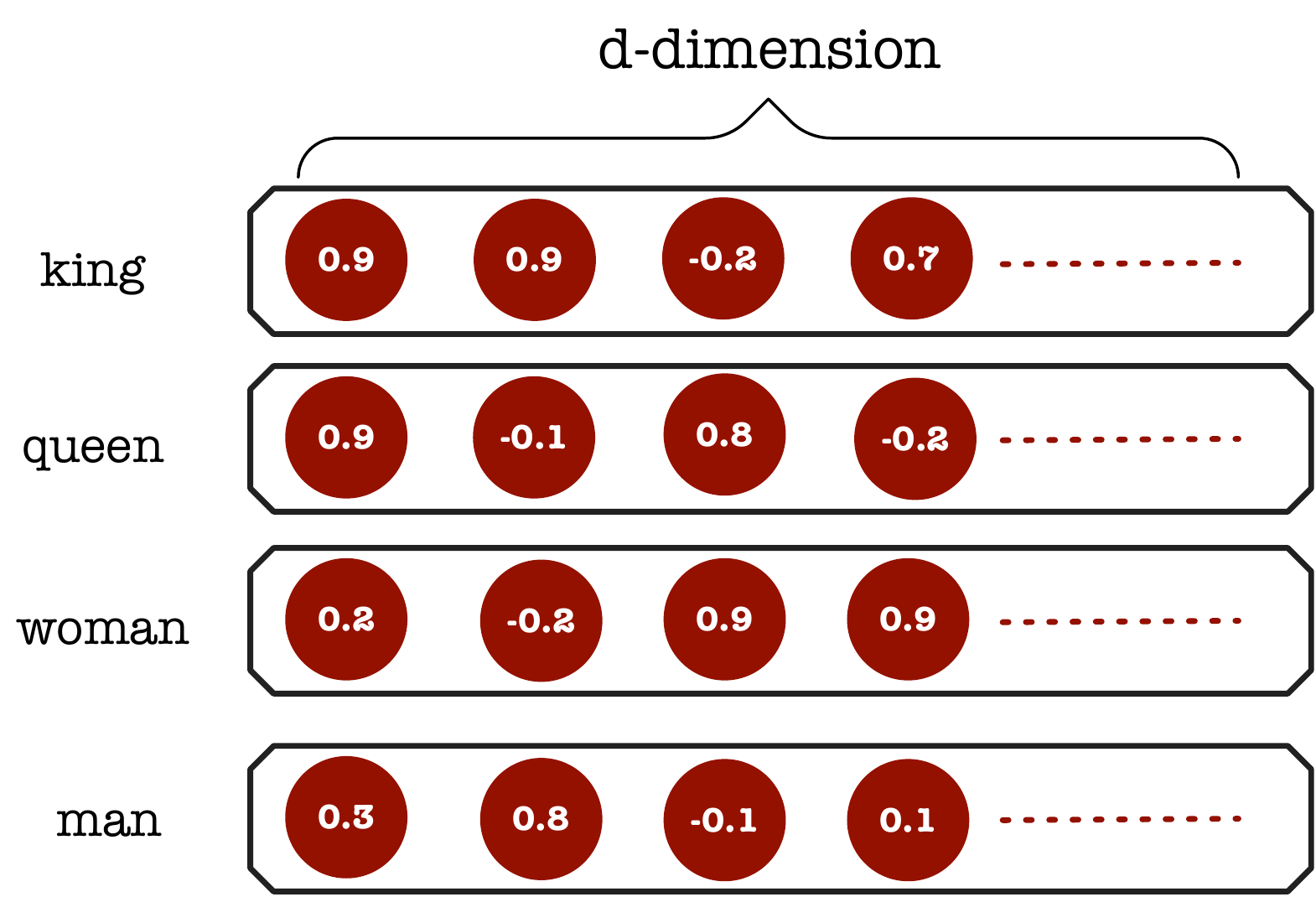}
     \vspace*{-2ex}
     \caption{Sample Word Embeddings}
     \label{fig:we}
     \vspace*{-3.5ex}
 \end{figure}

\stitle{Distributed Representation for Words.}
{\em \DRs of words} (\aka~{\em word embeddings}), recently introduced to deep learning,
are learned from the data such that semantically related words have embeddings that are often close to each other.
Typically, these approaches map each word in a dictionary into a high dimensional vector (\eg $300$ dimensions) 
where the geometric relation between the vectors of two words 
-- such as vector differencing or cosine similarity -- encodes a semantic relationship between them. 
There exists a cornucopia of \DRs for words including 
word2vec~\cite{mikolov2013distributed}, GloVe~\cite{pennington2014glove}, fastText~\cite{bojanowski2016enriching} and very recently ELMo~\cite{peters2018deep}.
Figure~\ref{fig:we} shows some sample word embeddings.
More discussion of \DRs for ER can be found in~\cite{deeper,anhaisigmod2018}.

\stitle{Distributed Representation for Tuples.}
We begin by treating each tuple as a document by breaking the attribute boundaries.
For example, a tuple with four attributes (title, author, venue, year) representing the foundational paper on database is represented as a document
``A relational model of data for large shared data banks. Edgar Codd. Communications of the ACM. 1970.''.

By treating each tuple as a sentence, one can leverage the extensive work from NLP 
that can encode sentences, paragraphs, and documents into a \DR~\cite{le2014distributed}.
However, these approaches are time consuming and recent works such as~\cite{wieting2015towards,arora2016simple} 
have shown that such complicated methods are often outperformed in a transfer learning setting by simpler compositional approaches.
Hence, we propose a simple, yet efficient and effective algorithm by synthesizing approaches 
from recent work on deep learning for ER~\cite{deeper,anhaisigmod2018} and theoretical underpinnings of word embeddings~\cite{arora2016simple,arora2018ALaCarte}.

\stitle{\DR Composition.}
Let us first consider a simple scenario where each of the words in the tuple is present in the \DR dictionary. 
A natural approach for composing a \DR for a tuple is to simply average the \DRs of all words in the sentence.
\vspace*{-2.5ex}
\begin{equation}
	\DR(t) = \frac{1}{|t|} \sum_{w \in t} \DR(w)
	\label{eq:tupleDRSimpleAveraging}
\end{equation}
\vspace*{-2ex}

$\DR(t)$ and $\DR(w)$ denote the \DRs for tuple $t$ and word $w$, respectively. 
However, it is possible to do better by computing the \emph{weighted} average of the \DRs instead of simple averaging.
Smooth Inverse Frequency (SIF)~\cite{arora2016simple} is one such weighting scheme that is reminiscent of TF-IDF weighting from IR.
Intuitively, the weight of a word is $a / (a + p(w))$ where $a$ is a parameter (typically set to $0.0001$) and $p(w)$ the frequency of word $w$.
This simple approach is known to outperform both simple averaging and even some complex approaches~\cite{arora2016simple,wieting2015towards}
due to the fact that the salient words often have a higher impact than common words.
The projection of the average vectors on their first principal component is then removed to reduce the bias.
The frequency of a word $w$ can be computed from the union of $D_S$ and $D_T$ 
or even from the union of all relations $\{D_S^1, \ldots, D_S^n\}$ from a given domain even if they are not involved in the given ER task.
\vspace*{-1ex}
\begin{equation}
	\DR(t) = \frac{1}{|t|} \sum_{w \in t} \frac{a}{a + p(w)} \DR(w)
	\label{eq:tupleDRSIFAveraging}
\end{equation}
\vspace*{-2.0ex}

\stitle{Handling Out-of-Vocabulary Words.}
A potential wrinkle in tuple composition happens when some of the words in the dataset are not present in the \DR dictionary.
This is dubbed as an out-of-vocabulary (OOV) scenario.
Of course, a simplistic and undesirable solution is to simply ignore such words when computing the \DR of a tuple 
or to use a random vector.
Prior work such as~\cite{deeper} also proposes expensive solutions such as building a domain specific \DR, 
using a probabilistic retrofitting approach, or subword information~\cite{bojanowski2016enriching}.

In this work, we use a simple technique that approximates the \DR of an unknown word from its context. 
Imagine that you are reading a sentence with a single unknown word $w$ with no access to a dictionary.
Usually, you will seek to infer the word's meaning from its context (a list of other \emph{known} words near $w$).
If you also have access to a list of other sentences where this unknown word $w$ occurs, 
you can get a better handle on the word's meaning by examining each of its contexts~\cite{arora2018ALaCarte}.
We empirically found that this approach provides a good-enough approximation of \DRs for the unknown words 
that is comparable with other approaches at a fraction of their time.
Formally, we compute \DR of $w$ as 
\vspace*{-1ex}
\begin{equation}
    \DR(w) = \frac{1}{|C_w|} \sum_{c \in C_w} \frac{1}{|c|} \sum_{w^{\prime} \in c} \DR(w^{\prime})
    \label{eq:oovALaCarte}
\end{equation}
\vspace*{-2ex}

\ni where $C_w$ is the set of contexts in which $w$ occurs. 
In our paper, we defined context to be $k=5$ words that occur before and after $w$. 

\stitle{Distributional Similarity Vectors.}
The next step is to compute the distributional similarity vector for a pair of tuples $t$ and $t^{\prime}$
that are passed as the input to the ML classifier. 
Popular approaches to compute the distributional similarity vector $\DR(t, t^{\prime})$ include 
vector differencing (op = subtraction) and Hadamard product (op = product).
In our paper, we use vector differencing. 
\vspace*{-1.5ex}
\begin{equation}
    \DR(t, t^{\prime}) = \sum_{i=1}^{d} |\DR(t)[i] \bullet \DR(t^{\prime})[i]|
    \label{eq:DRSimilarityVector}
\end{equation}

\stitle{Putting It All Together.}
Algorithm~\ref{alg:ERMetaFeatureStandardization} shows how to encode all the tuples in the same feature space 
such that their distributional similarity vector can be passed to an ML classifier for training/prediction under the Fellegi-Sunter model.

\begin{algorithm}[t!]
    \caption{Feature Space Standardization}
    \label{alg:ERMetaFeatureStandardization}
    {\small
    \begin{algorithmic}[1]
        \STATE \textbf{Input:} Target dataset $D_T$ 
		\STATE \textbf{Optional Input:} Source datasets $\mathcal{D}_S =\{D_S^1, \ldots, D_S^n\}$ 
        \STATE $W$ = Set of all distinct words from $D_T$
        \STATE $W_D$ = Set of all words from the \DR dictionary
        \STATE $W_U = W \setminus W_D$
        \STATE \add{/\!/ Compute \DR for unknown words}
        \FOR {each word $w \in W_U$}
            \STATE Collect context $C_w$ from $\mathcal{D}_S \cup D_T$
            \STATE Compute $\DR(w)$ using Equation~\ref{eq:oovALaCarte}
        \ENDFOR
        \STATE \add{/\!/ For computing weights for SIF}
        \STATE Estimate frequency of each word $w \in W$ over $\mathcal{D}_S \cup D_T$
        \FOR {each tuple $t \in D_T$}
            \STATE Compute $\DR(t)$ using Equation~\ref{eq:tupleDRSIFAveraging}
        \ENDFOR
        \STATE \add{/\!/ Computing distributional similarity vector}
        \FOR {each $(t, t^{\prime})$ in candidate set $C$ of $D_T$}
            \STATE $\DR(t, t^{\prime}) = |\DR(t) - \DR(t^{\prime})|$
        \ENDFOR
    \end{algorithmic}
    } 
\end{algorithm}

\stitle{Advantages of \DR based Feature Standardization.}

\sstab (1) All tuples from all datasets are encoded into the same feature space enabling reuse of ML classifiers.

\sstab(2) The feature space has a number of appealing properties such as encoding semantic similarity 
		and a fine-grained similarity computed holistically over $d=300$ dimensions.

\sstab(3) This standardization  allows us to pool training data from multiple source datasets -- 
        whereas under the feature truncation approach, this would have resulted in a small set of features that are present in each of the source datasets.

\sstab(4) It minimizes the effort of a domain expert in identifying appropriate features, similarity functions, and so on.

\sstab(5) Allows reuse of popular \DRs such as Word2vec, GloVe, and FastText that are trained on very large corpora. 

\sstab(6) Allows reuse of ML classifiers even from unrelated domains by leveraging the monotonicity of precision.

%% file: sec-algorithms.tex
\section{Algorithms}
\label{sec:algorithms}

Our next goal is to study methods of TL for ER.
After introducing some notations,
we will lay down different desiderata for devising TL algorithms for ER (Section~\ref{subsec:desiderata}).
We then describe our proposed algorithms (Section~\ref{subsec:algorithmcenarios}).




\stitle{Notation.}
For expository convenience, we abuse the use of
$D_S$ and $D_T$ as the sets of distributional similarity vectors of 
the source and target datasets $D_S$ and $D_T$ obtained through Algorithm~\ref{alg:ERMetaFeatureStandardization}. 
Let $D_S^L \subseteq D_S$ be the set of similarity vectors for which the labeled data is available. 
Let $D_S^U = D_S \setminus D_S^L$ be the corresponding set of unlabeled similarity vectors.
The sets $D_T^L, D_T^U$ are defined analogously for the target $D_T$.

\subsection{Algorithm Desiderata}
\label{subsec:desiderata}

There has been extensive prior work on TL algorithms for various domains~\cite{pan2010survey}.
However, most of these algorithms cannot be directly applied here as they are vulnerable to ER specific challenges or ignore key ER properties.
Our goal is to select and adapt an effective set of algorithms that perform well for ER. Most specifically, they must:

\sstab{\colorbox{black!25}{\bf D1:}}  
successfully address the ER-specific challenges such as imbalanced data, diverse schemata, and varying vocabulary,

\sstab{\colorbox{black!25}{\bf D2:}}  
be capable of leveraging key ER properties such as similarity vectors as features and monotonicity of precision,

\sstab{\colorbox{black!25}{\bf D3:}} 
be simple, efficient, and can be readily included in the existing ER pipelines such as Magellan~\cite{konda2016magellan},

\sstab{\colorbox{black!25}{\bf D4:}} 
work on classifiers that are widely used in ER,

\sstab{\colorbox{black!25}{\bf D5:}} 
be dataset and domain agnostic, and 
	
\sstab{\colorbox{black!25}{\bf D6:}} 
allow seamless transfer from multiple source datasets.
	

A key implication of these desiderata is that we would prefer algorithms that cause as little changes in the existing pipelines as possible.
In fact, all our algorithms require exactly one fundamental requirement -- \emph{the ability to set weights to each similarity vector}. 
All the major ML libraries from various languages support this functionality. 
This is the primary reason we prefer transferring data as against transferring ML classifiers.
While equally effective, the latter often requires complex modification of the objective function of the ML classifiers 
that are often non-trivial for a data cleaning expert and not available in the popular ER toolkits.

%
%
%

\subsection{TL Algorithms for ER Scenarios}
\label{subsec:algorithmcenarios}

We now describe our proposed algorithms for the different scenarios, which are based on the available training data in $D_S$ and  $D_T$ as introduced in Section~\ref{subsec:datascenarios}.

\begin{algorithm}[t!]
    \caption{{Scenario (Adequate, Nothing)}}
    \label{alg:instanceWeighting}
    {\small
    \begin{algorithmic}[1]
        \STATE \textbf{Input:} $D_S$ and $D_T$
		\STATE \add{/\!/ Determine weights of each $x_i \in D_S^L$.}
		\STATE $D = \{\}$
		\STATE $D = D \cup \{ \langle x_i, 1 \rangle \} \quad \quad \forall x_i \in D_S$
		\STATE $D = D \cup \{ \langle x_i, 0 \rangle \} \quad \quad \forall x_i \in D_T$
		\STATE Train a logistic regression classifier on $D$
		\STATE Create a weighted dataset $D_{w}$ where weight of $x_i \in D_S^L$ is computed from Equation~\ref{eq:weightingBickels}
		\STATE \add{/\!/ Train an ML classifier for ER for $D_T$}
		\STATE $M$ = ML classifier trained on $D_{w}$
		\STATE Apply $M$ on $D_T^U$ to detect duplicates
    \end{algorithmic}
    }
\end{algorithm}

\stitle{Scenario 1: (Adequate, Nothing).}
\label{subsec:AdequateNothing}
This is an extreme scenario where we do not have any training data $D_T^L$ from the target dataset.
However, we have an adequate amount of training data $D_S^L$ that was used to build a good ML classifier for $D_S$.
This scenario is called \emph{unsupervised domain adaptation} in TL.
At first blush, it might seem that we cannot do better than the naive transfer approach.
However, this ignores the availability of a key resource -- $D_T^U$.

Of course, $D_T^U$ cannot be directly used to train an ML classifier -- it does not contain any labeled data after all.
However, it might still be possible to indirectly use $D_T^U$ to improve the classifier trained on $D_S^L$.
Note that blindly training a classifier on $D_S^L$ \emph{optimizes} it for good performance on $D_S$ but not $D_T$.
This might be appropriate if $D_S$ and $D_T$ are extremely related -- otherwise
a better alternative is to ensure that the classifier is optimized for $D_T$ instead. 

The key insight is that the naive transfer approach assigns equal weights to the similarity vectors from $D_S^L$.
However, if we move towards a \emph{weighted paradigm} where different similarity vectors have different weights based on their fidelity to $D_T$, 
then the classifier would be able to achieve better performance on $D_T$. 
Suppose we have a mechanism to quantify the \emph{selection probability} that a similarity vector $x_i$ came from dataset $D_T$.
Then we must assign more weight to $x_i$ if the probability that it came from $D_T$ is high.
Meanwhile, we must assign low weights to the data points $x_i$ that have a low probability of coming from $D_T$.

There has been extensive work on different weighing schemes by ensuring that 
the  distance (such as KL-divergence~\cite{sugiyama2008direct} and kernel mean matching~\cite{huang2007correcting}) 
between the weighted $D_S^L$ and $D_T^U$ is minimized.
We advocate for an alternate weighing scheme originally proposed in~\cite{bickel2007discriminative} 
that can be easily incorporated into an existing ER pipeline.
The crucial observation is that an ML classifier can be used to generate the probability that $x_i$ came from a given dataset.

Our approach proceeds as follows.
We create a training dataset $D$ by combining $D_S$ and $D_T$.
We assign a label of $1$ to all similarity vectors in $D_S$ 
and a label of $0$ to all similarity vectors from $D_T$.
Note that this label is different from whether a given similarity vector in $D_S^L$ is a duplicate or not.
We then train an ML classifier (we use Logistic Regression in our paper) to predict whether a given similarity vector $x_i$ came from $D_S$ or $D_T$.
Specifically, we assign the following weight to each similarity vector $x_i \in D_S$.
\vspace*{-2ex}
\begin{equation}
	w(x_i) = \frac{1}{p(y=1|x_i)} - 1
	\label{eq:weightingBickels}
\end{equation}
\vspace*{-2ex}

We can see why this specific weighing scheme is appropriate for our purposes.
If the classifier can confidently predict $x_i$ as coming from $D_S$, then it is likely to be very different from $D_T$.
In such a case, we must assign a weight close to $0$. 
However, if the classifier is not very confident (say $p(y=1|x_i) \approx 0.5$), 
then it cannot clearly distinguish whether $x_i$ came from $D_S$ or $D_T$. 
This is a good candidate and must be assigned a higher weight close to $1$.
Algorithm~\ref{alg:instanceWeighting} describes the overall approach.


\begin{algorithm}[t!]
    \caption{{Scenario (Adequate, Limited)}}
    \label{alg:easyAdapt}
    {\small
    \begin{algorithmic}[1]
        \STATE \textbf{Input:} $D_S$ and $D_T$
		\STATE \add{/\!/ Feature augmentation}
		\STATE $D = \{\}$
		\STATE $D = D \cup \{ \langle \phi_S(x_i), y_i \rangle \} \quad \quad \forall (x_i, y_i) \in D_S^L$
		\STATE $D = D \cup \{ \langle \phi_T(x_i), y_i \rangle \} \quad \quad \forall (x_i, y_i) \in D_T^L$
		\STATE \add{/\!/ Train an ML classifier for ER on $D_T$}
		\STATE $M$ = ML classifier trained on $D$
		\STATE Apply $M$ on $D_T^U$ to detect duplicates
    \end{algorithmic}
    }
\end{algorithm}

\stitle{Scenario 2: (Adequate, Limited).}
\label{subsec:AdequateLimited}
The scenario where we have adequate labeled data $D_S^L$ and limited data from $D_T^L$
is called  \emph{supervised domain adaptation} in TL.
One could apply any of the three algorithms (NoT, NvT, and Algorithm~\ref{alg:easyAdapt}) for this scenario.
Our objective though is to design a classifier that achieves better performance by leveraging $D_T^L$.
Intuitively, if $D_S$ and $D_T$ are \emph{very} related, one could get good performance by simply \emph{pooling}
all data points from $D_S^L$ and $D_T^L$ and training a single classifier.
However, if they are not very related, a different approach is needed.

Consider a specific component $x^j$ of similarity vector $x$ that measures the similarity along dimension $j$.
If the datasets are not too related, then the importance of component $x^j$ might be different between $D_S$ and $D_T$.
If we train a classifier after blindly pooling $D_S^L$ and $D_T^L$, the learned parameters might not be appropriate.
However, we cannot only train on $D_T^L$ as it is limited.

The way out of this conundrum is \emph{feature augmentation} that enables us to 
learn parameters jointly when appropriate and learn individually otherwise.
There has been extensive work on feature augmentation such as~\cite{li2014learning,blitzer2006domain,kumar2010co}.
We advocate for a replication based feature augmentation~\cite{daume2007frustratingly,ciml} that requires minimal changes to the existing ER pipelines.
These approaches pre-process/transform datasets $D_S^L$ and $D_T^L$ 
that can then be passed to any ML classifier.

Given a similarity vector $x$ of dimension $d$, 
the approach in~\cite{daume2007frustratingly,ciml} 
creates a similarity vector of dimension $3 \times d$ by duplicating each feature $x^j$ in a clever manner that is different for $D_S$ and $D_T$.
\vspace*{-2ex}
\begin{align}
	\phi_S(x) = \langle x , x , \mathbf{0} \rangle \\	
	\phi_T(x) = \langle x , \mathbf{0}, x \rangle
\end{align}
\vspace*{-2ex}

The term $\mathbf{0}$ corresponds to a $d$-dimensional vector with all 0s.
The first $d$-dimensions correspond to common features while the next two $d$-dimensions are for source and target specific features respectively.
For example, let $x = [0.1, 0.9]$ be a similarity vector:
if $x \in D_S^L$, then $\phi_S(x) = [0.1, 0.9, 0.1, 0.9, 0, 0]$; and
if $x \in D_T^L$, then $\phi_T(x) = [0.1, 0.9, 0, 0, 0.1, 0.9]$.

This particular feature augmentation allows one to strategically share the parameters when necessary and learn individually otherwise.
The first $d$ dimensions allows us to learn shared parameters while the next $d$-dimensions allows us to learn parameters individually for $D_S$ and $D_T$, respectively.
We apply the appropriate transformation on each $x$ from $D_S^L$ and $D_T^L$ and feed the pooled transformed dataset to any ML classifier.
Algorithm~\ref{alg:easyAdapt} provides the pseudocode.


\begin{algorithm}[t!]
    \caption{{Scenario (Limited, Limited)}}
    \label{alg:easyAdaptPlusPlus}
    {\small
    \begin{algorithmic}[1]
        \STATE \textbf{Input:} $D_S$ and $D_T$
		\STATE \add{/\!/ Feature augmentation}
		\STATE $D = \{\}$
		\STATE $D = D \cup \{ \langle \phi_S(x_i), y_i \rangle \} \quad \quad \forall (x_i, y_i) \in D_S^L$
		\STATE $D = D \cup \{ \langle \phi_T(x_i), y_i \rangle \} \quad \quad \forall (x_i, y_i) \in D_T^L$
		\STATE \add{/\!/ Data + Feature augmentation}
		\STATE $D = D \cup \{ \langle  \phi_U(x_i), 0 \rangle \} \quad \quad \forall (x_i, y_i) \in D_T^U$
		\STATE $D = D \cup \{ \langle  \phi_U(x_i), 1 \rangle \} \quad \quad \forall (x_i, y_i) \in D_T^U$
		\STATE \add{/\!/ Train an ML classifier for ER for $D_T$}
		\STATE $M$ = ML classifier trained on $D$
		\STATE Apply $M$ on $D_T^U$ to detect duplicates
    \end{algorithmic}
    }
\end{algorithm}

\stitle{Scenario 3: (Limited, Limited).}
\label{subsec:LimitedLimited}
This is yet another tricky scenario where only limited amounts of labeled data $D_S^L$ and $D_T^L$ are available
that might not be sufficient for training effective individual classifiers for $D_S$ and $D_T$.
The objective here is to train an effective classifier for both $D_S$ and $D_T$ by pooling their respective labeled data.

Let us first dispense with an easier (sub) scenario.
If $D_S$ and $D_T$ are very related and the size of the pooled data $D_P = D_S^L \cup D_T^L$ is deemed to be adequate by a domain expert,
one can simply train an effective ML classifier by training it on $D_P$ or even using Algorithm~\ref{alg:easyAdapt}.

If $D_S$ and $D_T$ are not too related and/or pooled labeled data is still limited, 
then we need a more sophisticated approach that leverages the unlabeled data.
Without loss of generality, let us assume that we need to build an ML classifier for $D_T$.
We need to effectively pool not only $D_S^L$, $D_T^L$ but also $D_T^U$.
Note that if we wish to build an ML classifier for $D_S$, then we use $D_S^L$, $D_T^L$, and $D_S^U$.
This generic approach of effectively utilizing both the labeled and unlabeled datasets is known as {\em semi-supervised domain adaptation} in TL.

The key idea is to extend Algorithm~\ref{alg:easyAdapt} to make use of the unlabeled target data $D_T^U$.
Let us first consider what happens if we simply apply Algorithm~\ref{alg:easyAdapt} on $D_P$. 
If $D_P$ is not of adequate size, then the classifier will overfit $D_P$ and will not generalize well to $D_T^U$.
So, we need to force the learned classifier to do well on $D_T^U$ without having access to the labels.
An elegant approach to this problem was first proposed in~\cite{kumar2010co,daume2010frustratingly}.
First, we transform each $x \in D_T^U$ by 
\vspace*{-1ex}
\begin{equation}
\phi_U(x) = \langle 0, x , -x \rangle
\end{equation}
\vspace*{-2ex}

We perform \emph{data} augmentation by creating two copies of each $x \in D_T^U$.
We assign the label duplicate for one of them and non-duplicate to the other.
This seemingly bizarre augmentation is known to enforce a strong co-regularization on the classifier 
by ensuring that the weights learned for the transformed dataset also agree on the unlabeled data.
A rigorous analysis in terms of Rademacher complexity is provided in~\cite{kumar2010co}.
The pseudocode for this approach is given in Algorithm~\ref{alg:easyAdaptPlusPlus}.

\stitle{Scenario 4: (Adequate, Adequate).}
\label{subsec:AdequateAdequate}
This is a relatively straightforward scenario where we have adequate training data for $D_T$.
Any of the algorithms proposed in this section could be used here.

%% file: sec-putting-together.tex

%% file: sec-extensions.tex
\section{Practical Issues in TL for ER}
\label{sec:practicalIssues}

Next, we shift gears to discuss several issues that arise in practice and how to deal with them.
Section~\ref{subsec:TLMultipleDatasets} discusses TL for multiple sources.
Section~\ref{subsec:sourceDatasetSelection} describes the selection of source data, and
Section~\ref{subsec:modelEvaluation} addresses model selection.

\subsection{TL from Multiple Sources}
\label{subsec:TLMultipleDatasets}

Extending our algorithms to handle multiple source datasets is relatively straightforward.
Let $\{D_S^1, D_S^2, \ldots, D_S^n\}$ be the set of $n$ source datasets and $D_T$ be the target dataset.

\stitle{Algorithm NoT : No Transfer.}
This algorithm is not impacted by the number of source datasets as it simply trains from the target dataset.

\stitle{Algorithm NvT : Naive Transfer.}
We pool the similarity vectors from the labeled subset of all source datasets 
to create dataset $D_P$ and build ML classifier on it.

\stitle{Scenario 1: (Adequate, Nothing).}
Extension to multiple datasets requires two changes to Algorithm~\ref{alg:instanceWeighting}.
In Line~$4$, we assign a label of $1$ to similarity vectors from all source datasets $\{D_S^1, \ldots, D_S^n\}$.
As before, the similarity vectors from $D_T$ are assigned a label of $0$ and a logistic regression classifier is trained.
In Line~$7$, we compute the weights of the labeled similarity vectors from each of the source datasets using Equation~\ref{eq:weightingBickels}.
One can train any ML classifier on this weighted training dataset and apply it on $D_T^U$.

\stitle{Scenario 2: (Adequate, Limited).}
If there are $n$ source datasets with $d$ dimensions each, then the transformed feature space has $(n+2) \times d$ dimensions.
Given input $x$, the $i$-th $d$-dimensional window is set to $x$ for the transformation function $\phi_S^i$ of source $S_i$.
As before, the first and last $d$-dimensions are set to $x$ and $0$.
For example, if there $n=3$ datasets, the transformation functions are given by:
\vspace*{-1ex}
\begin{align}
	\phi_S^1(x) = \langle x , x , \mathbf{0}, \mathbf{0}, \mathbf{0} \rangle \qquad
	\phi_S^2(x) = \langle x , \mathbf{0} , x, \mathbf{0}, \mathbf{0} \rangle \\
	\phi_S^3(x) = \langle x , \mathbf{0} , \mathbf{0}, x, \mathbf{0} \rangle \qquad
	\phi_T(x) = \langle x , \mathbf{0} , \mathbf{0}, \mathbf{0}, x \rangle 
\end{align}
\vspace*{-3ex}

\stitle{Scenario 3: (Limited, Limited).}
The modification for this scenario is identical to that of Scenario 2, \ie $\phi_S$ and $\phi_T$.
The only additional change that is needed is the feature transformation function $\phi_U$ for unlabeled data.
Continuing the above example for $n=3$, we have:
\vspace*{-1ex}
\begin{equation}
	\phi_U(x) = \langle \mathbf{0}, \mathbf{0}, \mathbf{0}, x , -x \rangle
\end{equation}
\vspace*{-3ex}

\stitle{Scenario 4: (Adequate, Adequate).}
As mentioned before, any of the five proposed algorithms can be used for this scenario.
Depending on the specific algorithm, one can reuse the appropriate extension to multiple sources.

\subsection{Source Selection and Negative Transfer}
\label{subsec:sourceDatasetSelection}

The selection of an appropriate source dataset to transfer from
is of fundamental importance and has a disproportionate impact on the performance of the classifier on dataset $D_T$.
There has been extensive empirical research (such as \cite{rosenstein2005transfer}) that shows that 
a naive transfer between two dissimilar datasets can affect the performance of a classifier on the target dataset.
Negative transfer~\cite{perkins1992transfer,pan2010survey} is an extreme case where transfer from a source dataset
results in a \emph{reduced} performance on the targeted dataset.
However, research on dataset relatedness and negative transfer is still in its infancy thereby requiring guidance from a domain expert.

The key challenge is that dataset relatedness is inherently multidimensional in nature.
Two datasets might be different from each other in terms of 
prior probability distribution over labels or covariates, conditional distribution between labels and similarity vector,
sample selection bias, ratio/number of duplicates to non-duplicates, and many more.
So summarizing differences from all these dimensions into one quantitative measure is quite challenging.
We tackle two simpler sub-problems relevant to ER for which one can come up with effective heuristics.

\begin{algorithm}[t!]
    \caption{Source Relatedness via Selection Probability}
    \label{alg:srcRelatednessMCCAlgo}
    \begin{algorithmic}[1]
        \STATE \textbf{Input:} $D_S$ and $D_T$
		\STATE Add an origin column to $D_S$ and $D_T$ 
		\STATE Set origin column of $D_S$ to 1 and $D_T$ to 0.
		\FOR {iter = 1 to 10}
			\STATE $D$ = Randomly sample 80\% of $D_S$ and $D_T$
			\STATE Train classifier $M$ on $D$
			\STATE Evaluate performance of $M$ on $D_S \cup D_T \setminus D$ 
			\STATE Compute MCC using Equation~\ref{eq:mcc}
		\ENDFOR
		\IF {average MCC over 10 runs $\leq 0.2$}
			 \RETURN $D_S$ and $D_T$ are sufficiently related
		\ENDIF
    \end{algorithmic}
\end{algorithm}

\stitle{P1: Are $D_S$ and $D_T$ sufficiently related?}
The fundamental idea is to reuse the classifier from Scenario~1 of Section~\ref{subsec:AdequateNothing}, which we used to distinguish source and target datasets.
Intuitively, if the trained classifier has high accuracy, then the two datasets are very different and can be easily distinguished. 
The key insight is to reduce the notion of relatedness between two datasets 
to the selection probability of the aforementioned classifier and use a sophisticated statistical test for distinguishing them.
Let \at{TP, TN, FP, FN} be the number of true positives, true negatives, false positives, and false negatives of the classifier, respectively.
Matthews correlation coefficient (MCC)~\cite{baldi2000assessing} is defined as
\begin{equation}
	\at{MCC = \frac{TP \times TN - FP \times FN}{\sqrt{(TP+FP)(TP+FN)(TN+FP)(TN+FN)}}}
	\label{eq:mcc}
\end{equation}
Heuristically, a value of 0.2 or lower is considered as good~\cite{bigmlDatasetRelatedness}, \ie 
$D_S$ can be used as a source for $D_T$.
Of course, this statistical test has to be repeated many times for confidence.
Algorithm~\ref{alg:srcRelatednessMCCAlgo} shows the pseudocode.

\stitle{P2: Source Dataset Selection from a Pool of Sources.}
Another practical problem is to select one dataset $D_S$ from a set of candidate datasets $\{D_S^1, \ldots, D_S^n\}$.
One can use Algorithm~\ref{alg:srcRelatednessMCCAlgo} and rank the datasets based on MCC.
However, 
this may give incorrect ordering due to variance 
if two candidate datasets $D_S^i$ and $D_S^j$ are very similar to $D_T$.
We advocate for an alternate distance metric $d_{\mathcal{A}}$ proposed in~\cite{ben2007analysis,ciml}
that measures the distance between two distributions, $D_T$ and $D_S^i$,
based on how two different classifiers $f$ and $f^{\prime}$ can disagree on the labels from a given test set.
Note that $f$ and $f^{\prime}$ are chosen from the hypothesis class $\mathcal{H}$ that contains all classifiers of a given type 
(such as all SVMs or all logisitic regression classifiers).
It can be estimated~\cite{ben2007analysis} as:
\vspace*{-1ex}
\begin{equation}
	d_{\mathcal{A}} =2 \times (acc - 0.5)
	\label{eq:dAAccRelation}
\end{equation}
\vspace*{-2ex}

where $acc$ is the accuracy of the best classifier for domain separation.
Algorithm~\ref{alg:srcDatasetSelection} shows how one could use this metric to reliably select one or more sources.

\begin{algorithm}[t!]
    \caption{Source Dataset Selection from Candidates} 
    \label{alg:srcDatasetSelection}
    \begin{algorithmic}[1]
        \STATE \textbf{Input:} Source dataset candidates $\{D_S^1, D_S^2, \ldots, D_S^n\}$ and $D_T$
		\FOR {each dataset $D_S \in \{D_S^1, D_S^2, \ldots, D_S^n\}$}
			\FOR {iter = 1 to 10}
				\STATE $D$= Randomly select 80\% of data from $D_S$ and $D_T$
				\STATE Train classifier $M$ on $D$
				\STATE Compute held out accuracy of $M$ on $D_S \cup D_T \setminus D$
			\ENDFOR
			\STATE Compute average held out accuracy of $D_S$ and use it to estimate $d_{\mathcal{A}}$ using Equation~\ref{eq:dAAccRelation}
		\ENDFOR
		\RETURN dataset $D_S \in \{D_S^1, D_S^2, \ldots, D_S^n\}$ with least $d_{\mathcal{A}}$
    \end{algorithmic}
\end{algorithm}

\subsection{Model Evaluation with Limited Budget}
\label{subsec:modelEvaluation}

The performance of a trained ML classifier is evaluated on a separate test set.
However, in all but one scenario that we considered, 
we do not have sufficient labeled data $D_T^L$ for even training the classifier.
We need to avail the service of a domain expert subject to a labeling budget for estimating classifier's accuracy.
Note that this is distinct from active learning that uses a limited budget to \emph{train} a classifier.

We propose an effective heuristic to generate estimates of a classifier accuracy by leveraging the concept of stratified sampling~\cite{wasserman2013all}.
We fed a distributional similarity vector $t \in D_T^U$ to the classifier $M$ and obtained a probabilistic score $p(t)$. 
Given the desired number of strata $W$ from the expert (such as 5), we split the probability range $[0, 1]$ into $W$ equal sized partitions.
Each $t \in D_T^U$ is then assigned to its strata based on $p(t)$. 
For each strata $P_i$, we select $B_i$ samples uniformly at random,
get the true label from the domain expert and compute the classifier performance for $P_i$.
If the strata are sufficiently homogeneous, then this is a good estimate of the per-strata classifier accuracy. 

The next question is how to allocate the budget $B$ across different strata. 
Two intuitive options are equal allocation where $B_i = B/W$, and proportional allocation where $B_i = B \times \frac{|P_i|}{n}$. 
We recommend Neyman allocation~\cite{wasserman2013all} where the budget is allocated based on intra-strata variability $v_i$ as: 
\vspace*{-1ex}
\begin{equation}
    B_i = B \times \frac{v_i \times |P_i|}{\sum_{j=1}^{W} v_j \times |P_j|}
    \label{eq:perStrataNeymanAllocation}
\end{equation}
\vspace*{-2ex}

Each $t \in P_i$ is a Bernoulli random variable with success probability $p(t)$ whose variance is $p(t) \times (1-p(t))$. 
The sum of such all $t \in P_i$ follows a Poisson binomial distribution whose variance $v_i$ is approximated as:
\vspace*{-1ex}
\begin{equation}
	v_i = \sum_{t \in P_i} p(t) \times (1-p(t))
	\label{eq:varianceCalc}
\end{equation}
\vspace*{-3ex}

\begin{algorithm}[t!]
    \caption{Estimating Classifier Performance} 
    \label{alg:estimateClassifierAccuracy}
    \begin{algorithmic}[1]
        \STATE \textbf{Input:} Classifier $M$, Budget $B$, \#Stratas $W$ and $D_T^U$
        \STATE Stratify each $t \in D_T^U$ based on probability score $p(t)$ 
        \STATE Estimate $v_i$ and $B_i$ using Equations~\ref{eq:varianceCalc} and~\ref{eq:perStrataNeymanAllocation}
		\FOR {i = 1 to $W$}
            \STATE Random select $B_i$ samples from $P_i$
            \STATE Get labels from expert 
            \STATE Compute strata accuracy $A_i$ and recall $R_i$
		\ENDFOR
        \STATE Estimate dataset accuracy $A = \sum_{i=1}^{W} \frac{|P_i|}{n} A_i$
        \STATE Estimate dataset recall $R = \sum_{i=1}^{W} \frac{|P_i|}{n} R_i$
		\STATE Estimate dataset F-score from $A$ and $R$
    \end{algorithmic}
\end{algorithm}



%% file: sec-expt.tex
\section{Experiments}
\label{sec:expt}


In our evaluation, we answer the following key questions:
(i) {\em Feasibility of TL for ER:} 
Can we really improve the performance of popular ML classifiers for ER through TL?
(Exp-1);
(ii) {\em Effectiveness for Various Scenarios:}
What is the effectiveness of our proposed algorithms for the four scenarios discussed in Section~\ref{subsec:datascenarios}?
(Exp-2);
(iii) {\em Multi-Source TL for ER:}
Can we leverage multiple sources for TL?
(Exp-3);
(iv) {\em Cross Domain Transfer:}
What happens if we do not have enough training data from the same or similar domain?
(Exp-4);
(v) {\em Negative Transfer:}
Can we get decreased performance when applying TL for ER?
(Exp-5).

\subsection{Experimental Setup}
\label{subsec:expSetup}

\stitle{Hardware and Platform.}
All our experiments were performed on a quad-core 2.2 GHz machine with 16 GB of RAM. 
The algorithms were implemented in Python.
Scikit-Learn~\cite{scikit-learn} (version 0.19.1) was used to train the ML models.
We used Magellan~\cite{konda2016magellan} (version 0.3.1), an end-to-end 
ML-based ER framework for handling other ER stages 
such as feature generation, blocking, and evaluation.

\begin{table*}[t]
	\begin{center}
		\caption{Datasets for Experiments}
		\label{tbl:datasets}
		{\small
		\begin{tabular}{|c|c|c|c|c|c|}
			\hline
			{\bf Domain} & {\bf Dataset} & {\bf Notation} & {\bf \#Tuples} & {\bf \#Duplicate Pairs} & {\bf \#Attributes}\\ \hline
			\multirow{3}{*}{Publications}  & DBLP-ACM~\cite{erhardws} & Pub-DA & 2,616 - 2,294 & 2,224 & 4 \\ \cline{2-6}
			& DBLP-Scholar~\cite{erhardws} & Pub-DS & 2,616 - 64,263 & 5,347 & 4 \\ \cline{2-6}
			& Cora~\cite{synthesizer} & Pub-C & 993-993 & 14, 280 & 9 \\ \hline
			\multirow{3}{*}{Movies} & IMDB-OMDB~\cite{magellandata}& Mv-IO & 1,132,262-2,300,984& 521,590& 11\\ \cline{2-6}
			& DBPedia-IMDB~\cite{tkdeBlockingFramework}  & Movies-DI & 27,615-23,182 & 22,405 & 5 \\ \cline{2-6}
			& Rotten Tomatoes-IMDB~\cite{magellandata} & Movies-RI & 7,390-6,407 & 190 & 10 \\ \hline
			\multirow{2}{*}{Songs} & Million Song Data-1~\cite{bertin2011million} & MSD-1 & 15,233-15,233 & 10,000 & 7 \\ \cline{2-6}
			& Million Song Data-2~\cite{bertin2011million} & MSD-2 & 14,966-14,966 & 10,000 & 7 \\ \hline
			\multirow{2}{*}{Restaurants} &Yellow Pages-Yelp~\cite{magellandata} & Rest-YY1 & 11,840-5,223 & 130 & 6 \\ \cline{2-6}
			& Yelp-Yellow Pages~\cite{magellandata} & Rest-YY2 & 9,947-28,787 & 88 & 12\\ \hline
			\multirow{2}{*}{Books} & Amazon-Barnes \& Noble~\cite{magellandata} & Books-AN & 9,836-9,958 & 99 & 7 \\ \cline{2-6}
			& Barnes \& Noble-Half~\cite{magellandata} & Books-BH & 3,022-3,099 & 327 & 8 \\ \hline
		\end{tabular}
		} 
	\end{center}
\end{table*}

\stitle{Datasets.}
In order to highlight the wide applicability of our proposed approaches, 
we conducted extensive experiments over 12 datasets from 5 diverse domains, namely publications, movies, songs, restaurants, and books (See Table~\ref{tbl:datasets}  for details).
The datasets from the publication, movie, and song domains are popular benchmark datasets 
which have been used in prior ER work with both ML- and non-ML-based approaches.
The datasets from the restaurant and book domains 
are from the Magellan data repository~\cite{magellandata}.
Their ground truth duplicates were only partially provided.
We trained multiple ML models on the partial data and predicted the match status for the remaining tuple pairs using an Ensemble ML classifier from Magellan.
The output was then manually verified and corrected. 

As the name suggests, the Million Song Dataset (MSD) has 1 Million songs.
Our proposed algorithms were quite effective over this dataset.
We chose two subsets MSD-1 and MSD-2 in an adversarial manner such that they
have a large $d_{\mathcal{A}}$. 
This reduces the effectiveness of TL making this a challenging dataset. 

\stitle{Distributed Representations.}
We used FastText~\cite{bojanowski2016enriching} as our default \DR that was trained over 16 billion tokens from Wikipedia 2017 and other related web corpus.
Unless otherwise specified, we used the default dimension of $d=300$.
We leveraged the re-training functionality of FastText to partially avoid the out-of-vocabulary scenario.
Specifically, we pooled all the datasets from a given domain into one giant document where each tuple corresponds to a sentence.
We trained FastText on this document with the pre-trained \DRs from Wikipedia as initial values.
This re-training provides better initialization for unknown words from the datasets we used for ER. 
We initialized the \DRs of rare words -- defined as those that are not in FastText's dictionary and occurring less than 5 times in our domain --
using the method described in Section~\ref{subsec:drForWordsTuples}~\cite{arora2018ALaCarte}.
We empirically found that this provides better initialization than the subword based mechanism of FastText.
We used Equation~\ref{eq:tupleDRSIFAveraging} to convert tuples to \DRs.

\stitle{Algorithms.}
We evaluated the five algorithms
two of which are baselines - No Transfer~(NoT) and Naive Transfer~(NvT).
The other three correspond to the scenarios (Adequate, Nothing),  (Adequate, Limited), and  (Limited, Limited).
When the available training data is less than a 10\%, we consider it to be limited (\ie Scenario 2 or 3).

\stitle{ML Classifiers.}
For our experiments, we primarily focus on four ML classifiers:
Logistic Regression (LR), Support Vector Machines (SVM), Decision Trees (DT), and Random Forests (RF).
These classifiers are widely used for ER in practice, extensively evaluated in prior work, and supported by most ER frameworks.
Note that our approaches are quite generic and can support many popular ML classifiers.

\stitle{TL for ER Best Practices.}
Both TL and ER suffer from imbalance issues separately.
ER has a class imbalance issue where the number of non-duplicates are many orders of magnitude larger than the number of duplicates.
TL often has data imbalance where the amount of training data for source $D_S^L$ is often much larger than that of target data $D_T^L$.
Individually, these issues are readily addressable.
When combined, they require a careful solution that we describe below.

A typical solution for class imbalance in ER is to under sample the non-duplicates such that 
the ratio between the number of duplicates and non-duplicates is less than a threshold.
We used a ratio of 1:3 whenever feasible.
As an example, let the training data has 100 duplicates and 10,000 non-duplicates.
Suppose that we wish to train a classifier with 200 tuple pairs. 
Then, we will pick 50 duplicates and 150 non-duplicates -- uniformly at random -- such that 1:3 ratio is maintained.
When we wish to train with 1,000 tuple pairs, we pick all the available 100 duplicates and pick 900 non-duplicates uniformly at random.
We then apply the balanced weighting heuristic proposed in~\cite{king2001logistic} that assigns different \emph{class weights} to duplicates and non-duplicates.

The dataset imbalance problem in TL can be solved analogously.
Consider a scenario where $|D_S^L|=1000$ while $|D_T^L|=100$.
On the one hand, if we blindly use this for training, the source dataset can ``swamp'' the target dataset.
On the other hand, due to the data scarcity, we cannot afford to ignore $900$ tuple pairs from $D_S^L$ to achieve dataset parity.
One effective approach is to apply different \emph{instance weights} to the training data.
For example, each $t \in D_S^L$ gets a weight of $0.1$ while each $t \in D_T^L$ gets a weight of 1.
While very effective, this approach cannot always be used in our paper.
Recall that in the algorithm for scenario (Adequate, Nothing), we use the 
class separator ML classifier to get the weights for each tuple in $D_S^L$.
In this case, we use the technique of probabilistic \emph{replication} 
whereby we repeatedly sample with replacement from $D_T^L$ till we get 1,000 samples. 
Of course, one can also use the even simpler replication where each $t \in D_T^L$ is replicated 10 times in total.

A related complication arises where there are constraints on the number of tuples one can use for training in $D_S^L$ and $D_T^L$.
For example, in our experiments we vary the size of $D_S^L$ and $D_T^L$ to understand their respective impact.
This also often occurs in practice where the source data is much larger than the target data.
Suppose we have $|D_S^L|=100,000$ and $|D_T^L|=100$.
Clearly, utilizing all the data (even with a weight of $0.01$) might swamp the target dataset.
For our experiments, we fixed the maximum data imbalance ratio as 10.
Hence, we require at most 1,000 tuples from $D_S^L$ when $|D_T^L|=100$.
We use the following heuristic : we use Algorithm~\ref{alg:instanceWeighting} to obtain weight for each $t \in D_S^L$ 
and perform importance sampling on $D_S^L$ with these weights.
We found that blindly picking the top-1,000 tuples with the largest weights is not very robust.
This approach was also used for selecting tuples from $D_S^L$ for Algorithms~\ref{alg:easyAdapt} and~\ref{alg:easyAdaptPlusPlus}.
In keeping with known TL best practices (\eg Chapter~8 of~\cite{ciml}),
the weights were computed in both cases using Equation~\ref{eq:weightingBickels} \emph{before applying transformations} $\phi_S, \phi_T$ and $\phi_U$.
Once again, we perform importance sampling to pick the necessary number of tuples (such as 1,000 in this example)
and then apply the transformations on them.

\subsection{Experimental Results}
\label{subsec:expResults}

\begin{figure}[t!]
         \flushleft
         \mbox{
                 \subfigure[Pub-DA]{\includegraphics[width=125pt]{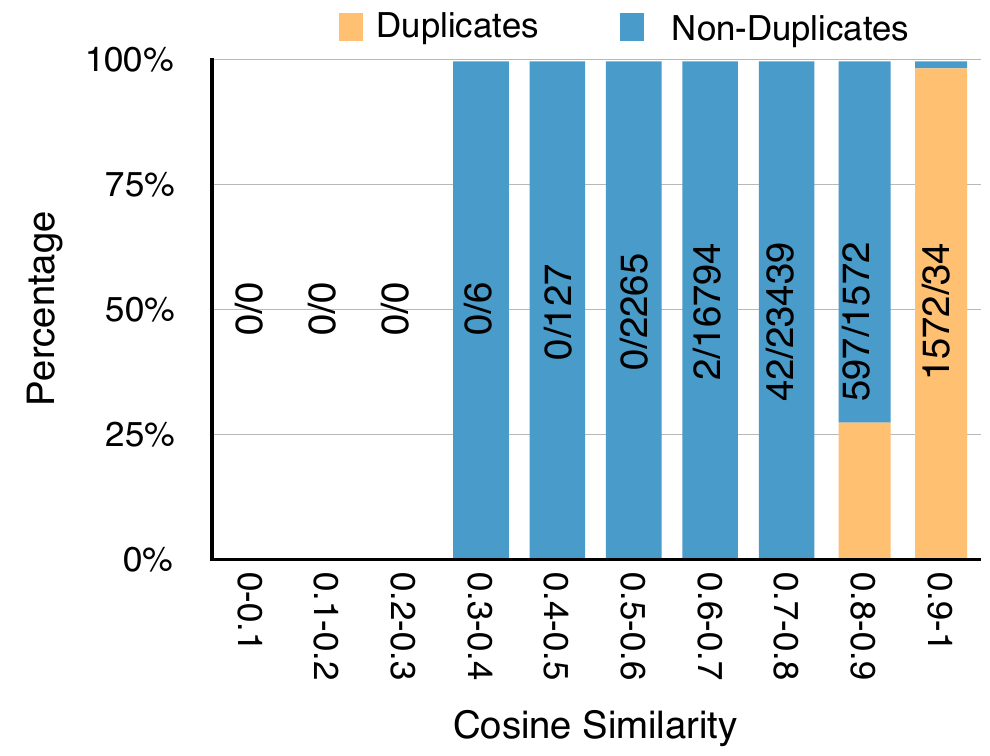}\label{subfig:pub-da}}
                 \subfigure[Pub-DS]{\includegraphics[width=125pt]{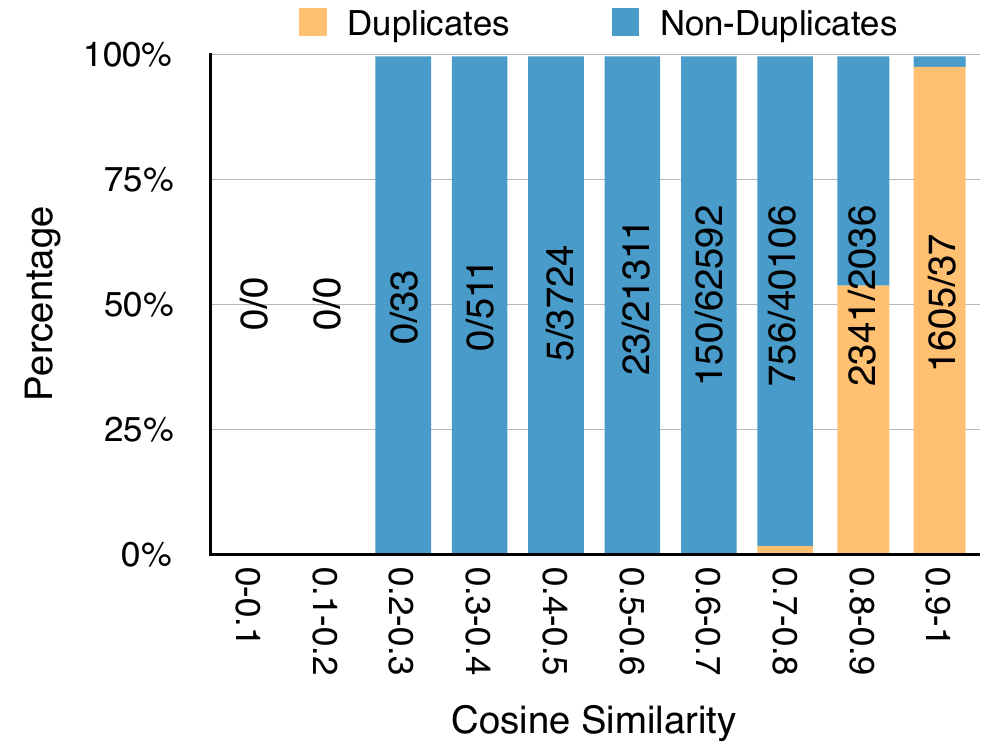}\label{subfig:pub-ds}}
         }
         \caption{Challenges of Duplicate Distributions}
         \label{fig:tlForERChallenges}
 \end{figure}

\stitle{Exp-1: TL for ER Challenges.}
Recall from Section~\ref{sec:introduction} that TL for ER is challenging due to a number of issues such as 
class/data imbalance and shift in prior/class conditional probability.
Figure~\ref{fig:tlForERChallenges} visually highlights some of these challenges and also the potential for transfer.
This figure shows the class based histogram of distributional similarity of two datasets - Pub-DA and Pub-DS.
Consider Figure~\ref{fig:tlForERChallenges}(a) corresponding to dataset Pub-DA.
For each distinct tuple pair $(t_i, t_j)$ in the blocked candidate set, we construct their respective \DRs 
and compute their Cosine similarity~\cite{manning2017representations}.
The non-duplicates and duplicates are highlighted in different colors.
We can see that the histograms of these two datasets are sufficiently similar to plausibly allow transfer
while sufficiently different to make such transfer challenging.

\etitle{Understanding Baseline Algorithms.}
We first compare the performance of the baseline algorithms No Transfer (NoT) and Naive Transfer (NvT).
Even though we refer to them as baselines, they often provide formidable performance in most ER tasks.
This is due to three reasons.
Using the Fellegi-Sunter paradigm, the input to all ML classifiers for various datasets is a distributional similarity vector of fixed dimension.
This allows a straightforward transfer opportunity.
Second, the \DR-based approach allows one to measure holistic similarity in a high dimensional ($d=300$) vector space.
Third, when combined with the monotonicity of precision property, we can see that 
a distributional similarity vector with a high similarity is very likely to be a duplicate regardless of which dataset it was computed from. 
So an ML classifier can readily identify high similarity distributional vectors as a duplicate without much (re)training!

Let us consider the performance of the algorithms in two datasets - (Pub-DA, Pub-C) and (MSD-1, MSD-2).
Given two datasets $D$ and $D^{\prime}$, we present the (source, target) order such that it produces the worse result.
In other words, if $(D, D^{\prime})$ has better result than $(D^{\prime}, D)$, we present the latter.
We measure the F-measure of four classifiers while varying the size of the training data used.
For NoT, this measures the size of $D_T^L$ while it measures size of $D_S^L$ for NvT.

 \begin{figure}[t!p]
         \centering
         \mbox{
                 \subfigure[(Pub-DA, Pub-C)]{\includegraphics[width=120pt]{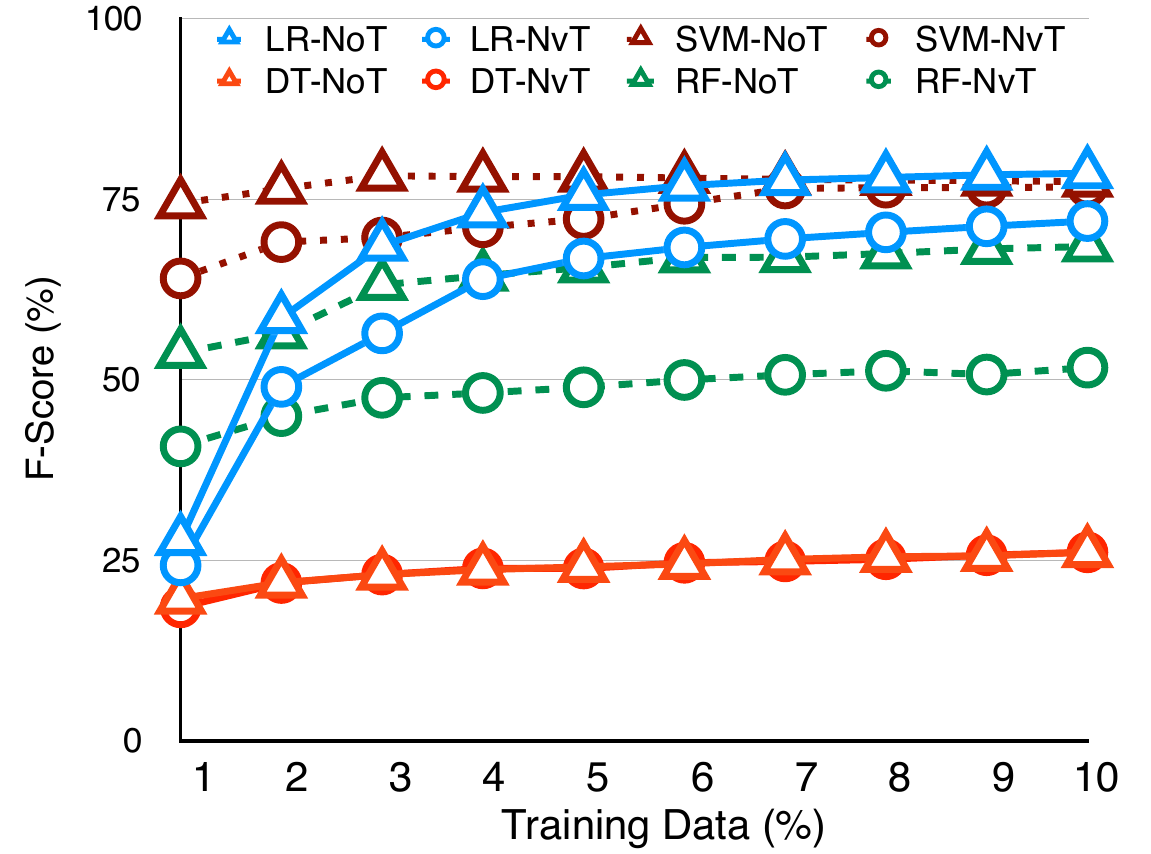}\label{subfig:baseline-pub}}
                 \subfigure[(MSD-1, MSD-2)]{\includegraphics[width=120pt]{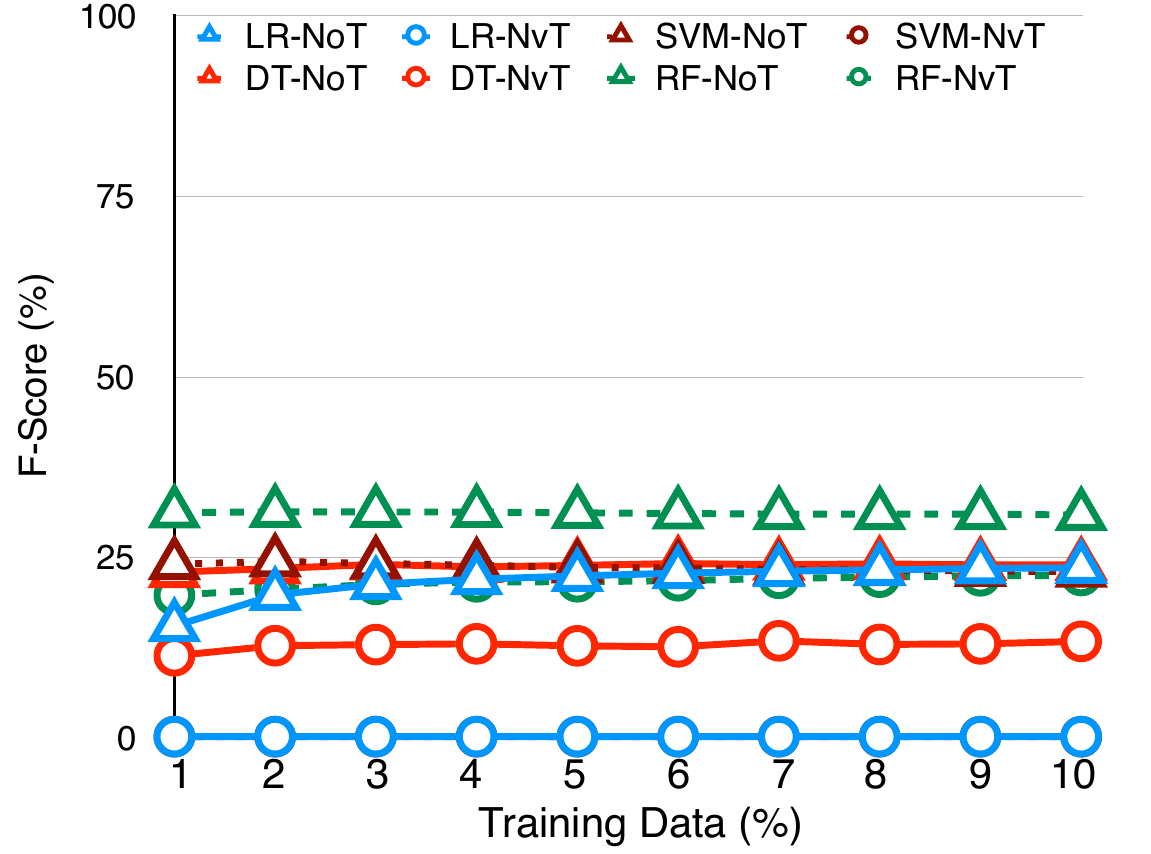}\label{subfig:baseline-msd}}
         }
      \vspace*{-4ex}         
         \caption{Behavior of Baselines}
         \label{fig:baselines}
 \end{figure}
 
\addtolength{\subfigcapskip}{-0.1in}
\begin{figure*}[t!p]
         \centering
         \mbox{
                 \subfigure[Publications]{\includegraphics[width=125pt]{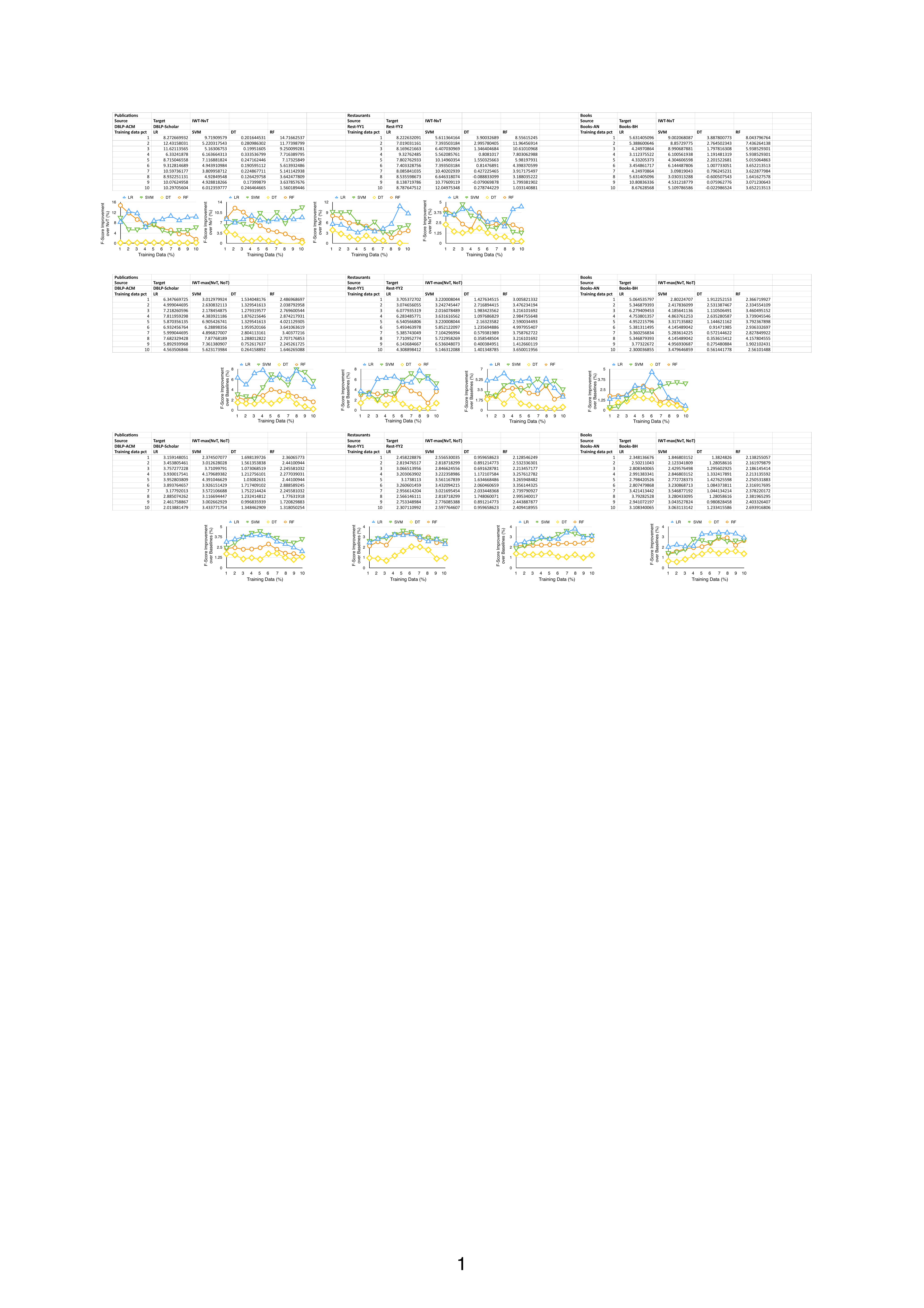}\label{subfig:sce1-pubs}}
                 \subfigure[Restaurant]{\includegraphics[width=125pt]{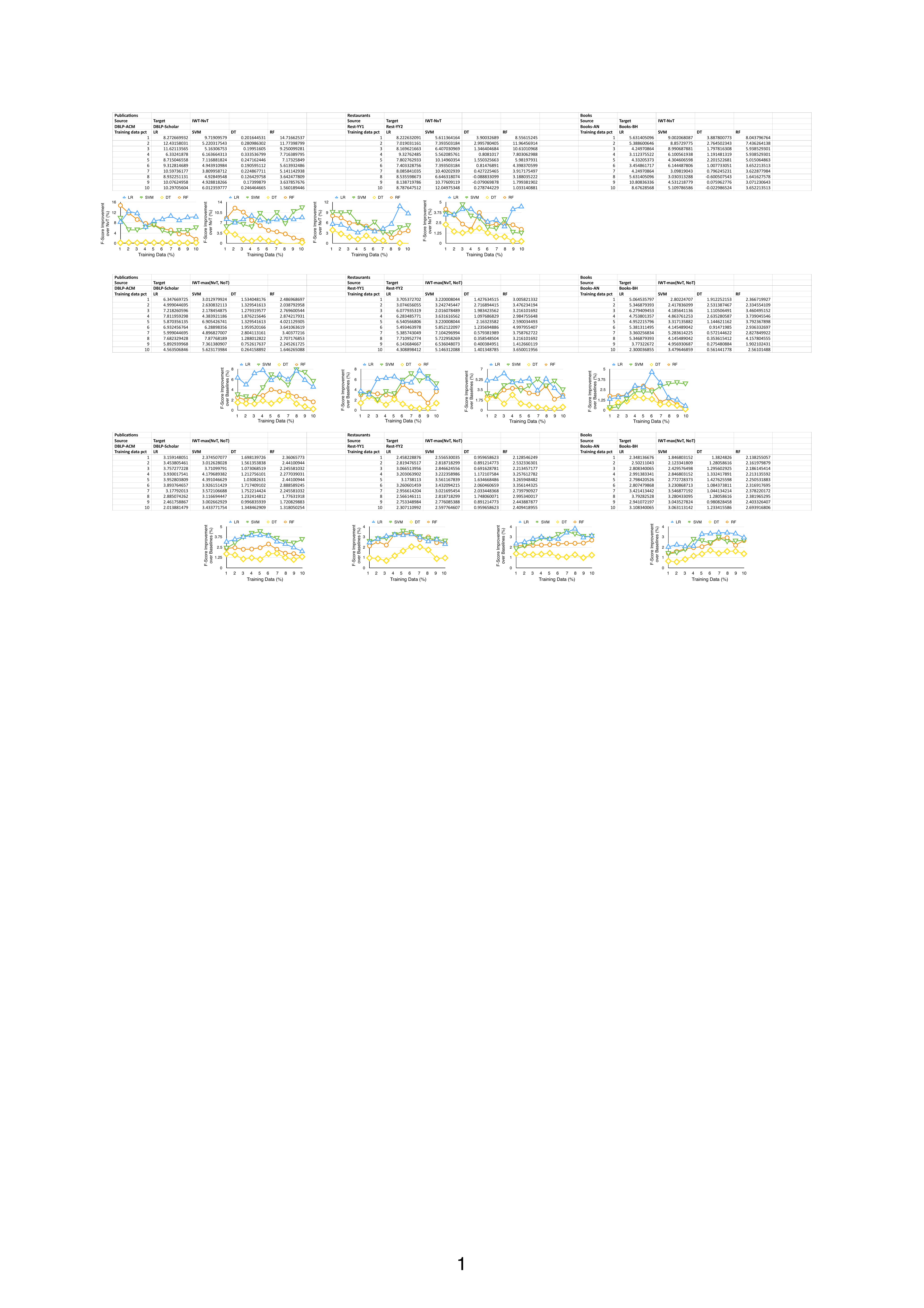}\label{subfig:sce1-restaurants}}
                 \subfigure[Books]{\includegraphics[width=125pt]{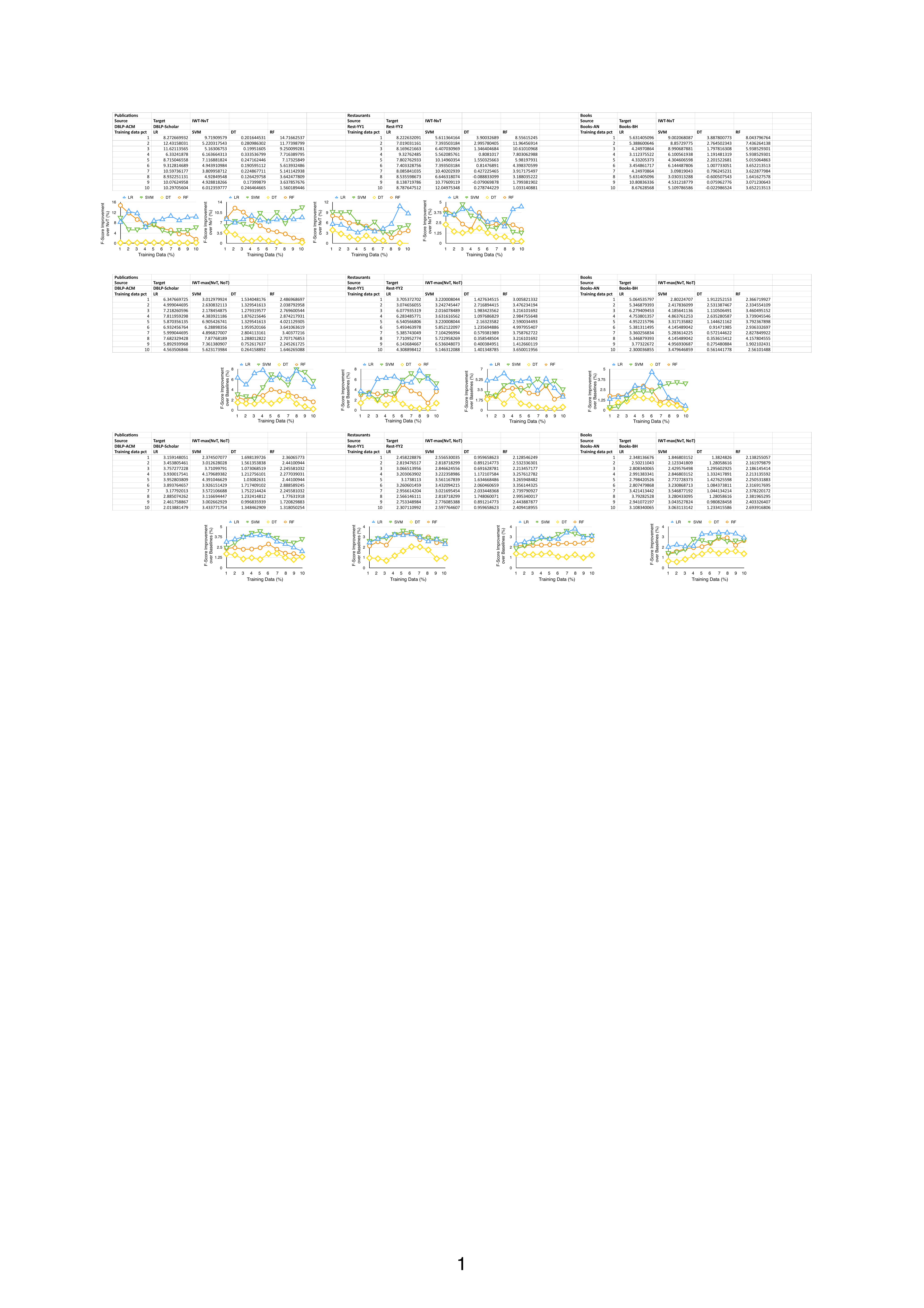}\label{subfig:sce1-books}}
                 \subfigure[Songs]{\includegraphics[width=125pt]{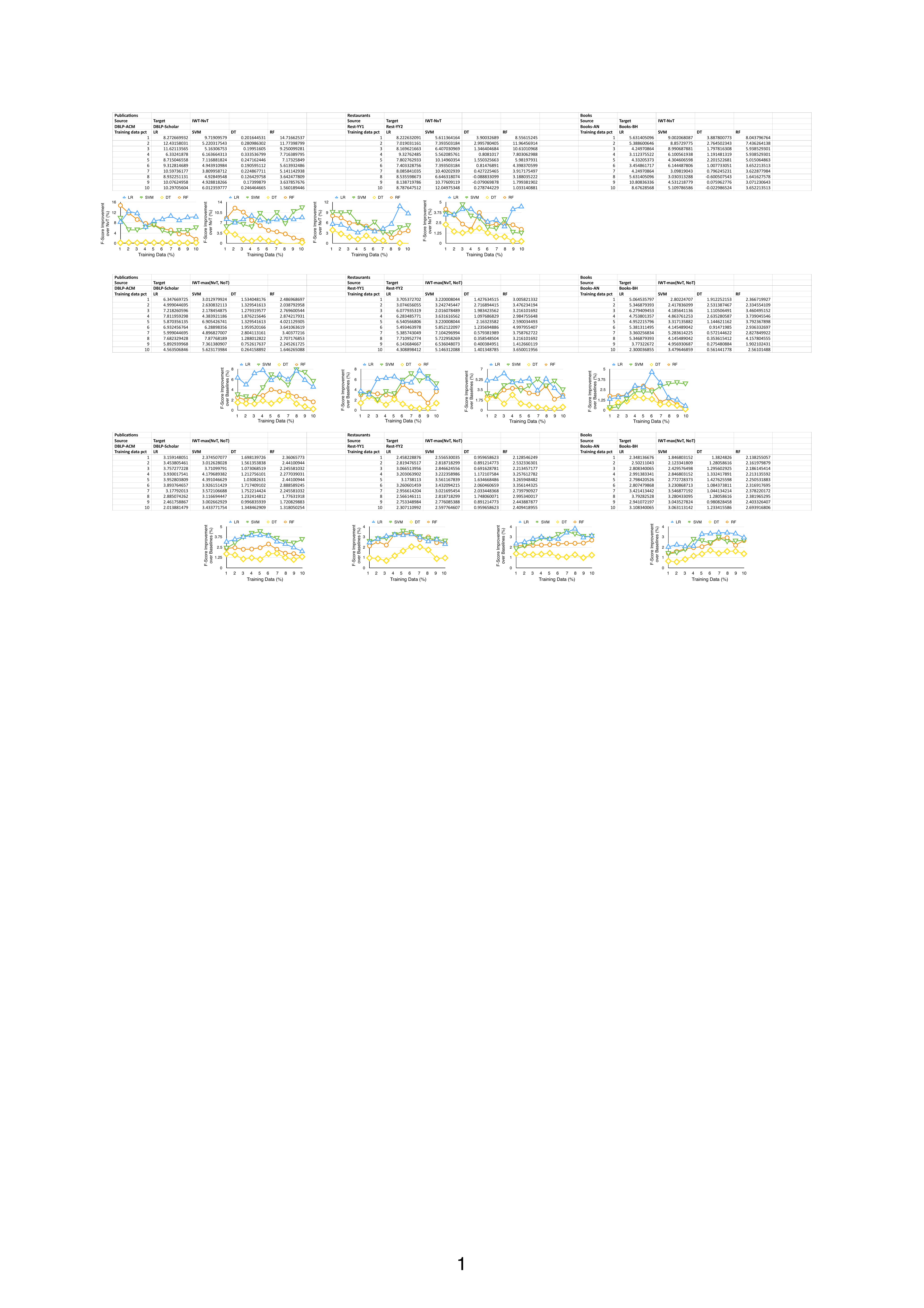}\label{subfig:sce1-songs}}
         }
      \vspace*{-4ex}         
         \caption{Scenario 1. (Source, Target): (Adequate, Nothing)}
         \label{fig:scenario1}
 \end{figure*}
 \addtolength{\subfigcapskip}{0.1in}
 
\addtolength{\subfigcapskip}{-0.1in} 
  \begin{figure*}[t!p]
         \centering
         \mbox{
                 \subfigure[Publications]{\includegraphics[width=125pt]{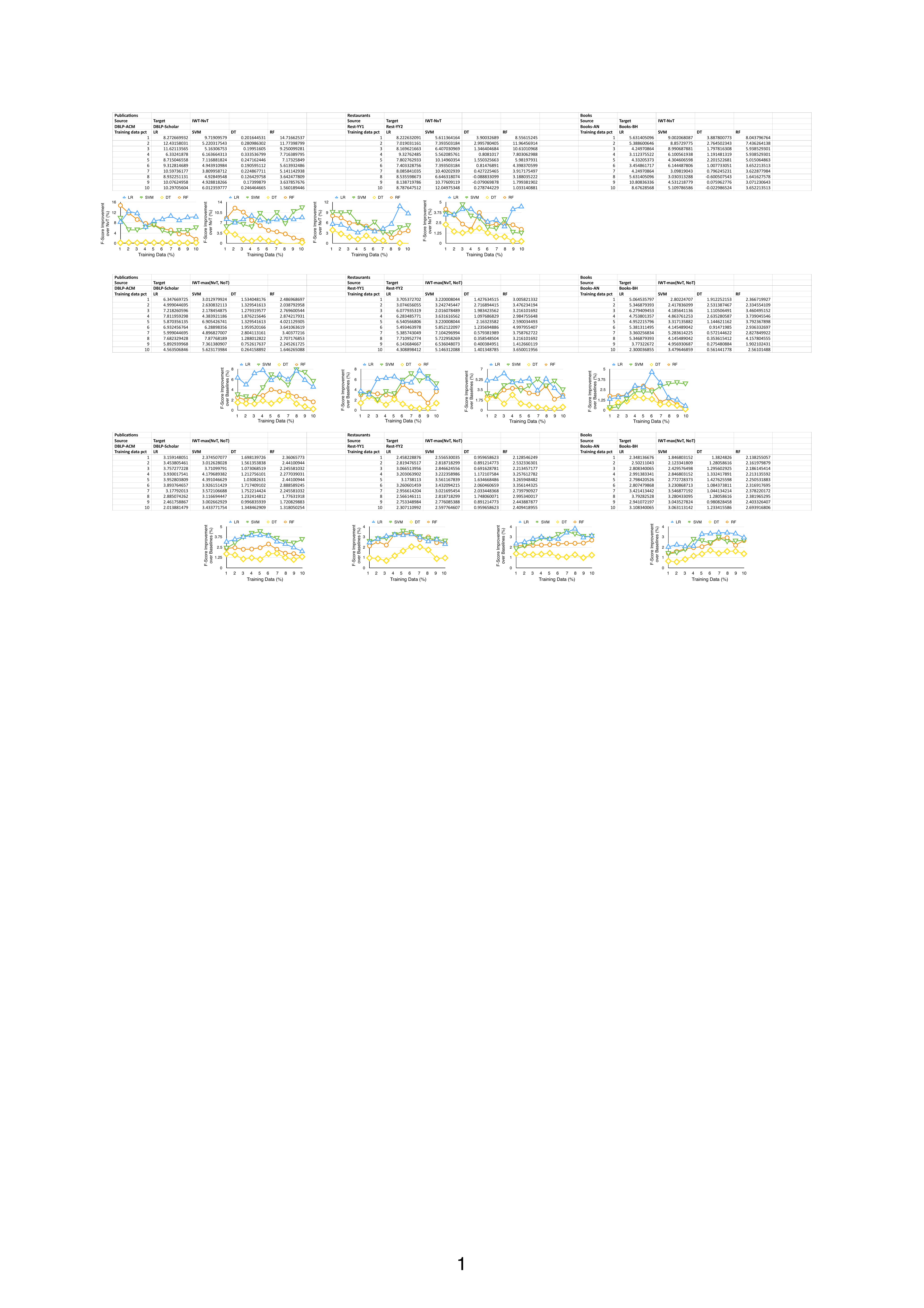}\label{subfig:sce2-pubs}}
                 \subfigure[Restaurant]{\includegraphics[width=125pt]{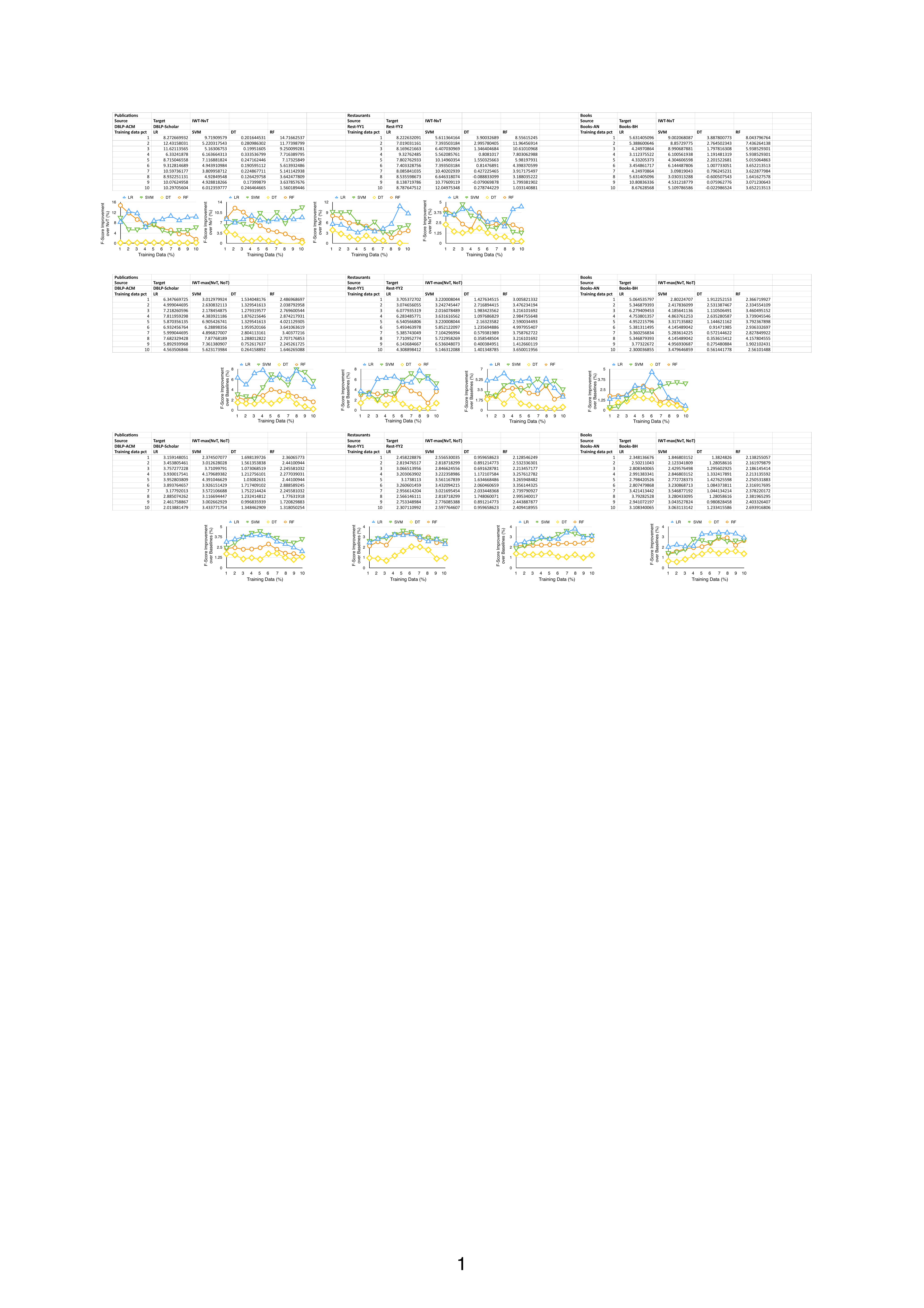}\label{subfig:sce2-restaurants}}
                 \subfigure[Books]{\includegraphics[width=125pt]{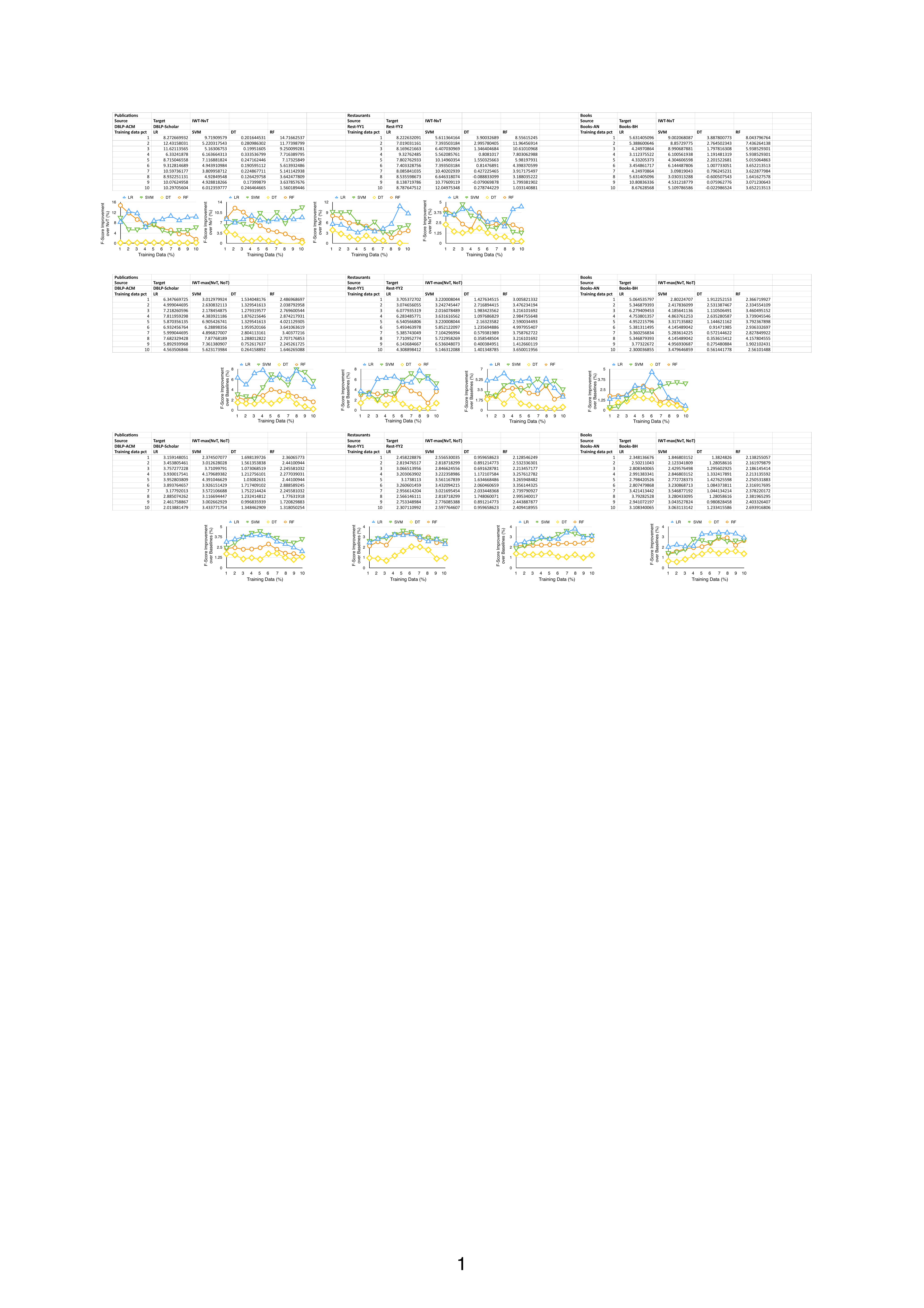}\label{subfig:sce2-books}}
                 \subfigure[Songs]{\includegraphics[width=125pt]{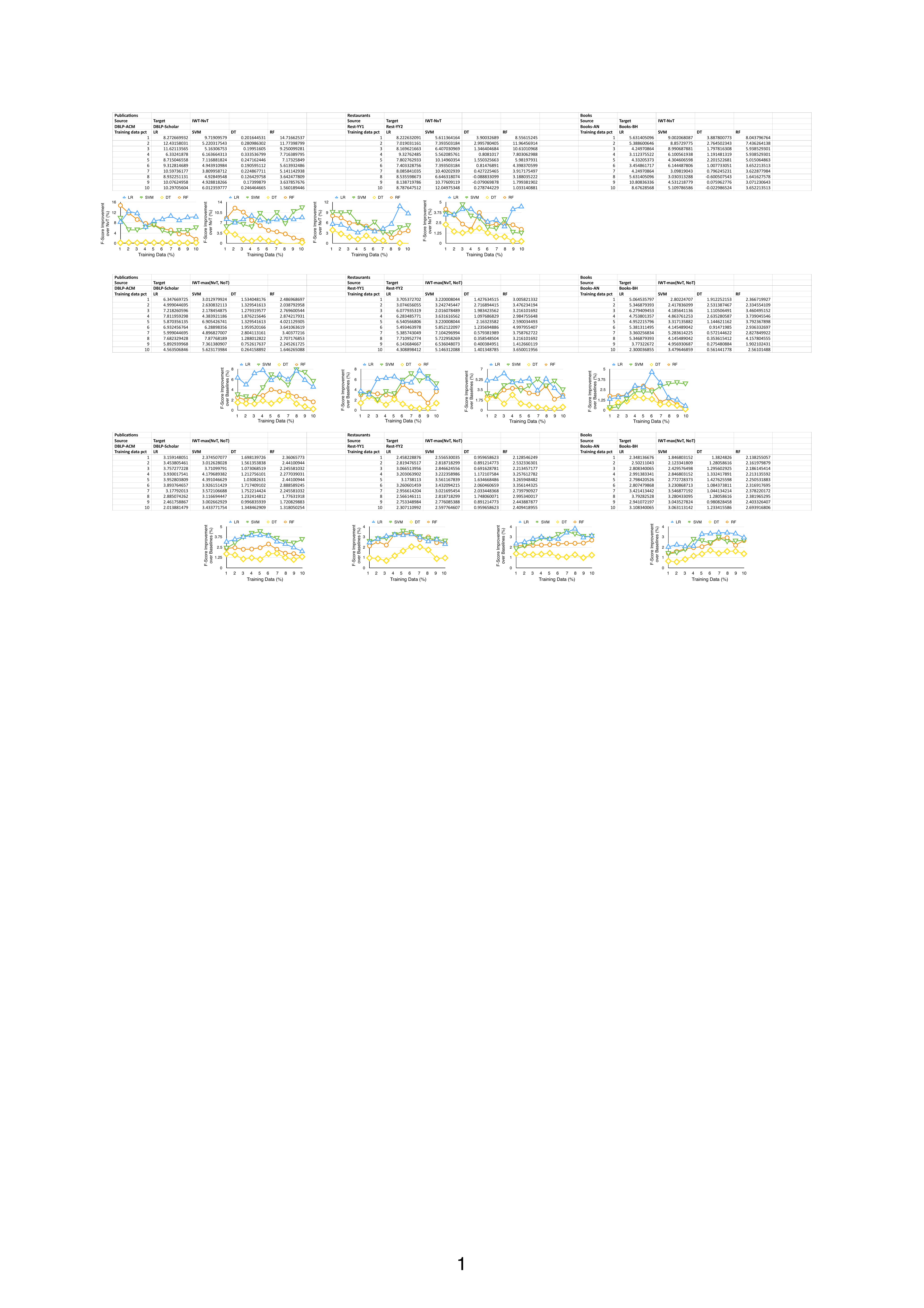}\label{subfig:sce2-songs}}
         }
      \vspace*{-4ex}         
         \caption{Scenario 2. (Source, Target): (Adequate, Limited)}
         \label{fig:scenario2}
 \end{figure*}
\addtolength{\subfigcapskip}{0.1in}
 
\addtolength{\subfigcapskip}{-0.1in}
  \begin{figure*}[t!p]
         \centering
         \mbox{
                 \subfigure[Publications]{\includegraphics[width=125pt]{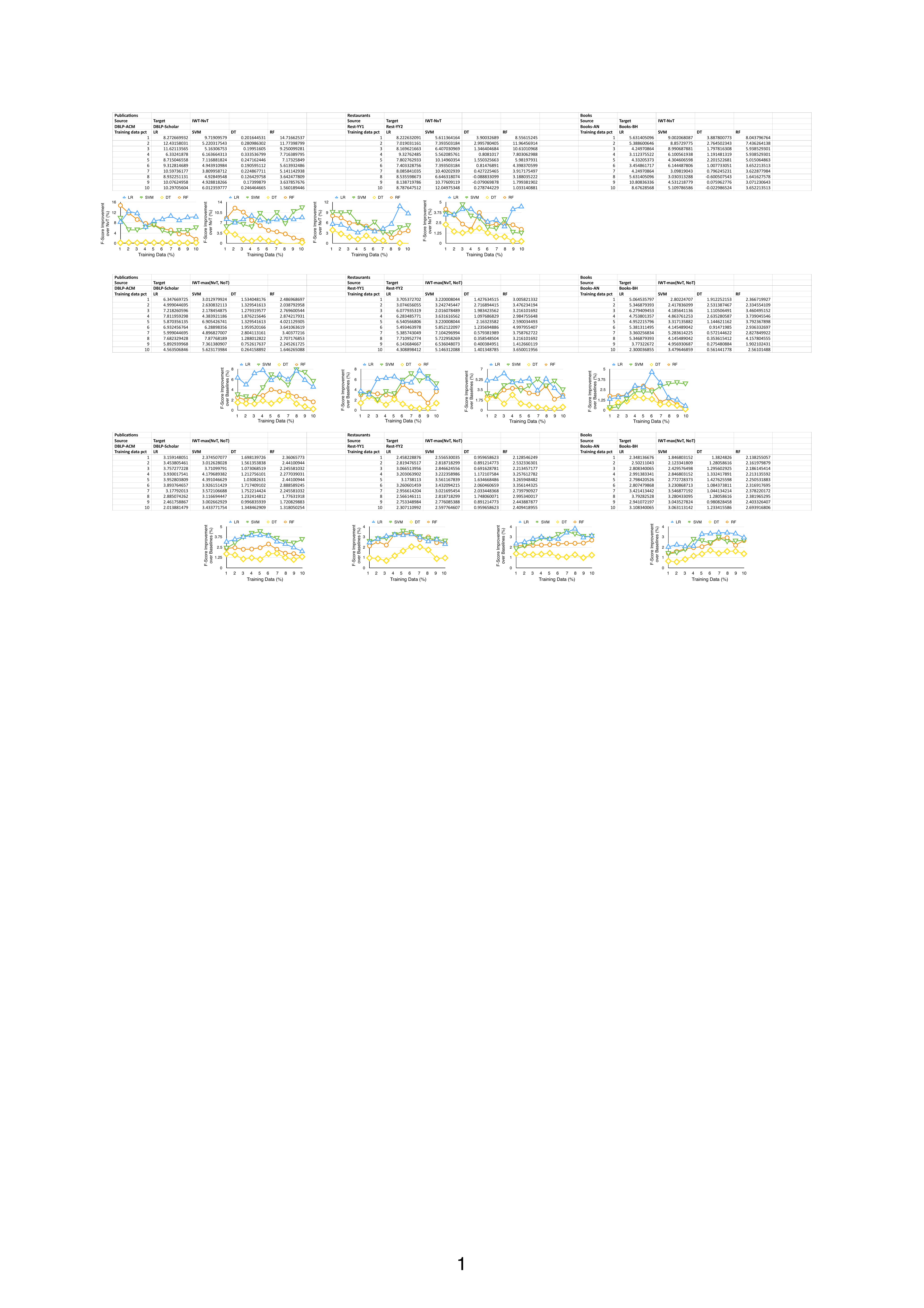}\label{subfig:sce3-pubs}}
                 \subfigure[Restaurant]{\includegraphics[width=125pt]{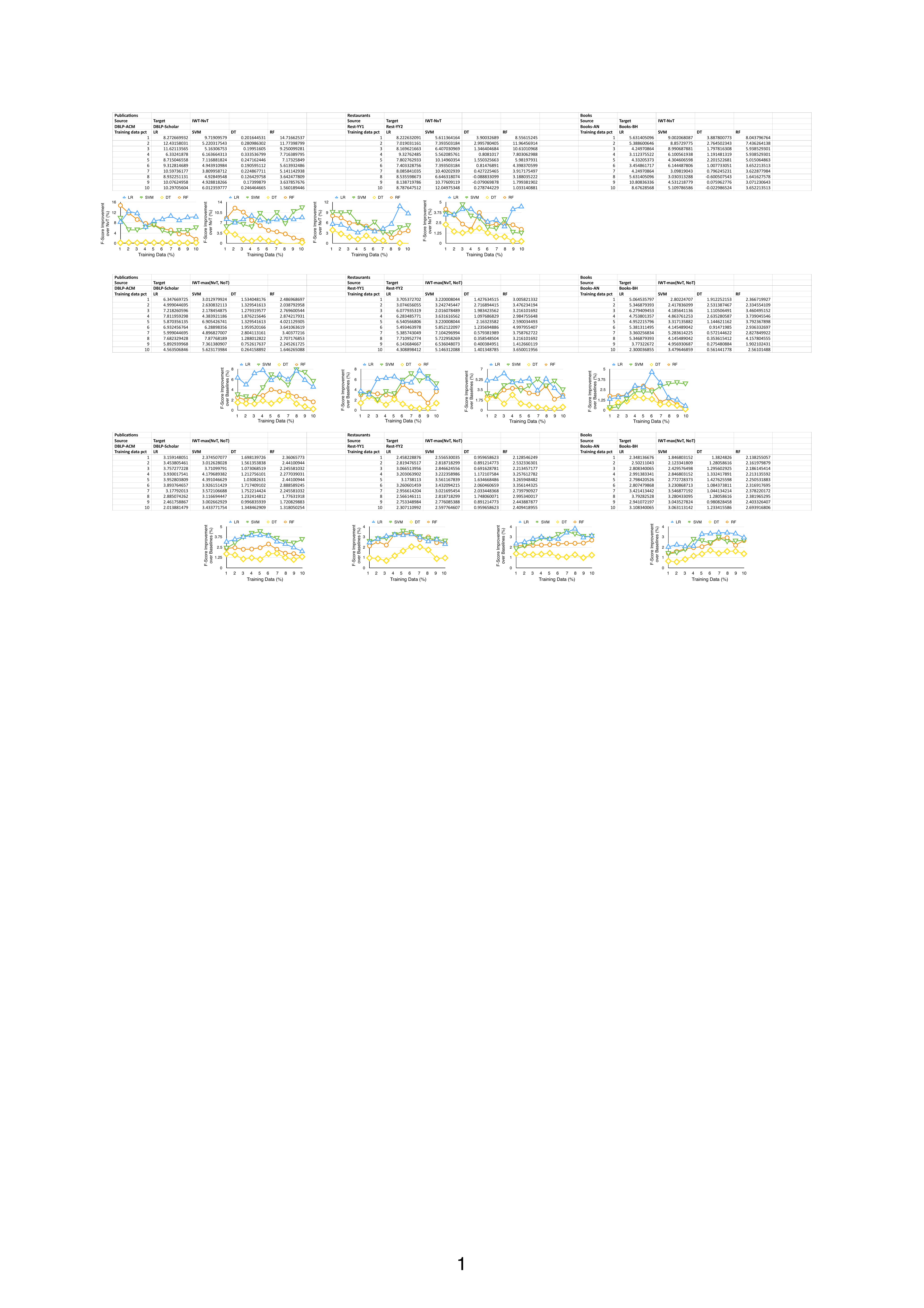}\label{subfig:sce3-restaurants}}
                 \subfigure[Books]{\includegraphics[width=125pt]{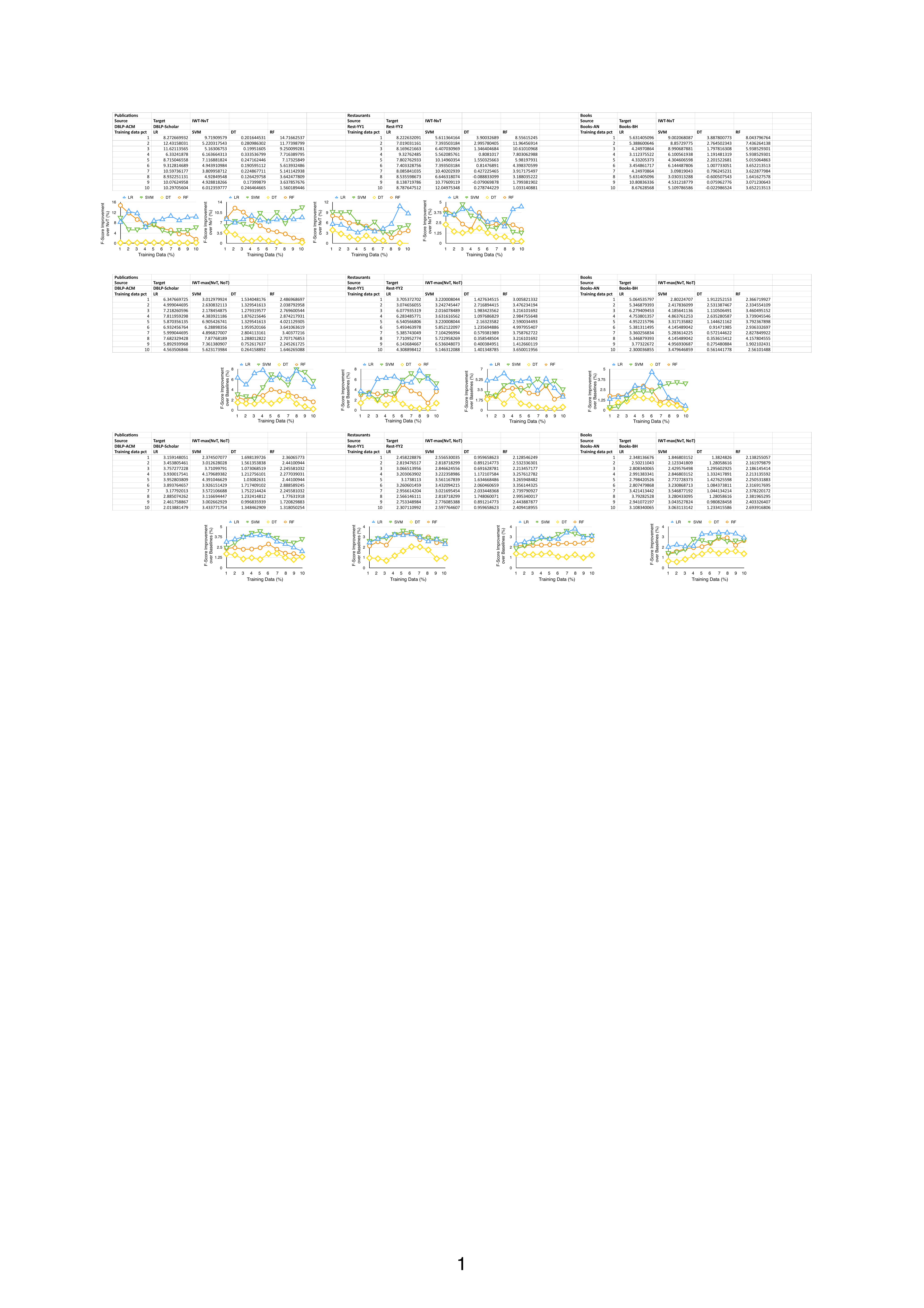}\label{subfig:sce3-books}}
                 \subfigure[Songs]{\includegraphics[width=125pt]{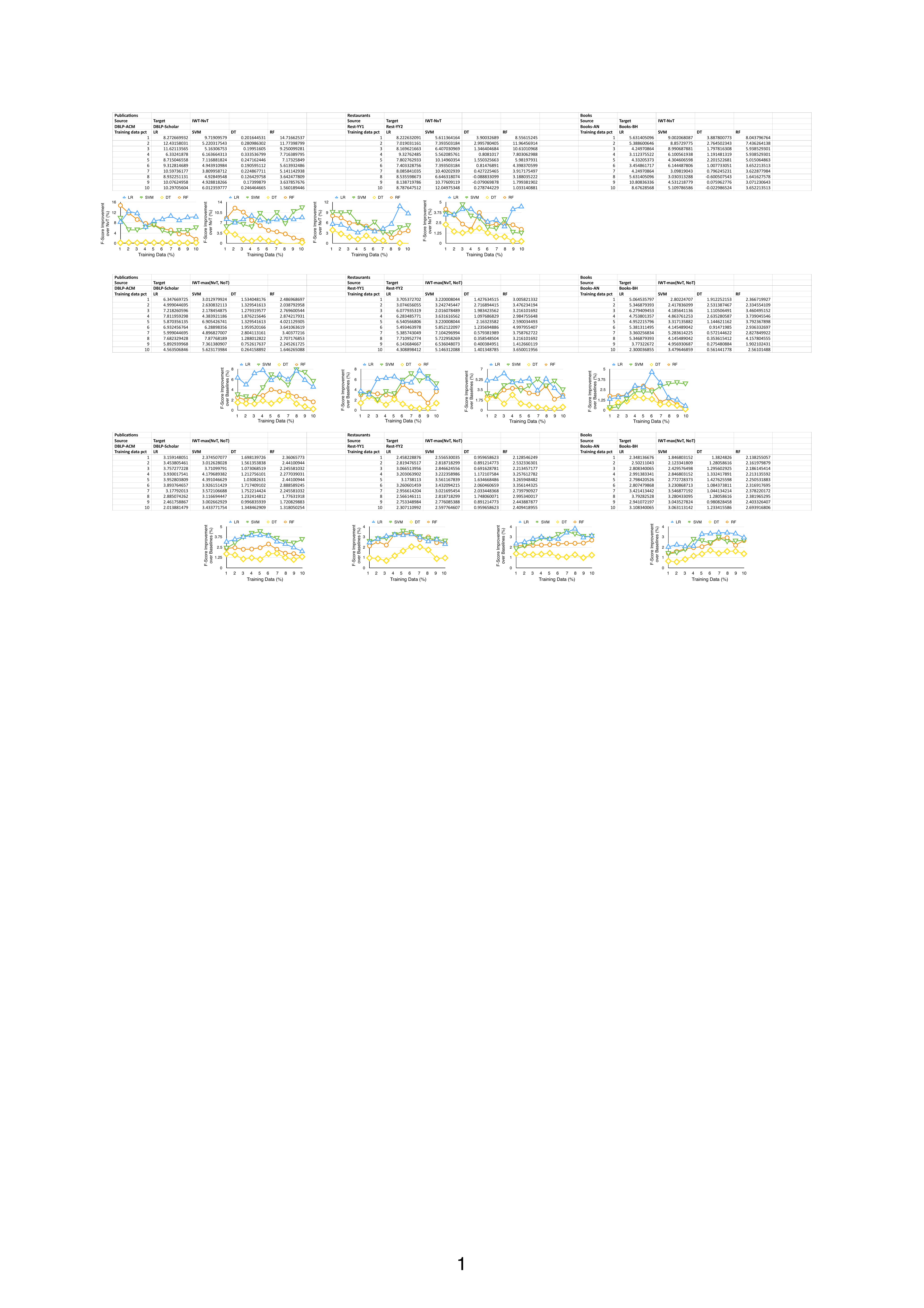}\label{subfig:sce3-songs}}

         }
      \vspace*{-2ex}         
         \caption{Scenario 3. (Source, Target): (Limited, Limited)}
         \label{fig:scenario3}
 \end{figure*}
\addtolength{\subfigcapskip}{0.1in}

Figures~\ref{subfig:baseline-pub} and~\ref{subfig:baseline-msd} show the results.
We can make a few observations that hold true for both datasets.
First, the baselines do provide strong performance and can be used when the sophisticated TL algorithms cannot be applied.
The performance of the baselines improve slightly with larger training data size.
Both NoT and NvT suffers from a plateauing where the performance improvement slows down as more and more $D_S^L$ is utilized.
This is to be expected as the two datasets have a number of differences such as data/class imbalance, shift in covariance, prior probability, and class conditional probability.
Second, as the size of training data increases, the comparatively complex ML models of RF, SVM, and LR outperform DT.
Overall, the baselines provide good results for (Pub-DA, Pub-DS).
However, the situation is flipped for (MSD-1, MSD-2).
Recall that this dataset was constructed in an adversarial manner.
Both SVM-NvT and LR-NvT provide F-scores close to 0 while 
the others  perform relatively better.
Even then, the F-score peaks at approximately 20\% regardless of the size of the training data.

\stitle{Exp-2: Evaluating TL's Impact on ER.}
Next we evaluate the improvement achieved by our algorithms over the baselines. 
We would like to bring attention to three key factors in order to appreciate the increased performance.
First, once the feature space of the datasets are standardized using our proposed approach, 
the baselines already provide substantial improvements.
This improvement happens consistently across multiple domains.
Second, we intentionally operate in a resource constrained environment where we
use as little as 1\% to as much as 10\% of the relevant training data 
from $D_S^L$ or $D_T^L$ (as explained in each scenario).
In fact, this setting both stress tests our algorithm and is also most relevant to 
real-world scenarios where very often only a small amount of labeled data is available.
Finally, our improvements in F-score for various datasets vary between 1\%-12\% with 3-5\% being the typical value.
It is important to note that ER is an extremely well studied problem and achieving such improvements is quite challenging!
In fact, even when a number of novel approaches for ER such as 
crowdsourcing~\cite{GokhaleDDNRSZ14,DBLP:journals/pvldb/WangKFF12} and deep learning~\cite{deeper} are used, 
the achieved improvements in F-score was often in single digits despite using substantially more training data.

\etitle{Scenario 1: (Adequate, Nothing).}
We evaluate our proposed approach for the scenario where we set $D_T^L = \emptyset$.
We vary the size of $D_S^L$ from 1\%, 2\% all the way to 10\%.
In order to reduce the clutter, we show the improvement in F-score obtained by Algorithm~\ref{alg:instanceWeighting} compared to NvT.
Figures~\ref{fig:scenario1} shows the results.
Overall, the performance across all the domains is very similar.
Our proposed approach consistently outperforms NvT - often by a significant margin.
The improvement in F-score becomes close to 0 for more than 10\% of training data.
The estimated weights of those additional tuples is close to 1 thereby making our approach similar to NvT. 
While this scenario cannot be solved using traditional ER approaches, our proposed approach not only makes it feasible but also provides a good performance.

\etitle{Scenario 2: (Adequate, Limited).}
In our next set of experiments, we evaluate the effectiveness of Algorithm~\ref{alg:easyAdapt} for Scenario 2.
We fix the size of $D_T^L$ to 10\% and vary the size of $D_S^L$.
To reduce clutter, we show the difference in F-score between our algorithm and the F-score of the \emph{best of the two baselines}. \ie NoT or NvT.
Figures~\ref{fig:scenario2} shows the results.
As expected, even with very small amount of $D_T^L$ and $D_S^L$, our TL approach outperforms both  baselines consistently. 

\etitle{Scenario 3: (Limited, Limited).}
We next evaluate the effectiveness of Algorithm~\ref{alg:easyAdaptPlusPlus} by varying the size of \emph{both} $D_S^L$ and $D_T^L$ from 1\% to 20\%. 
We fix the size of the unlabeled data $D_T^U$, which is needed by the algorithm, to 20\% of the training data.
Once again, the results show that our approach outperforms the baselines.
The difference in performance comes from the fact that training ML classifiers 
on small-ish training data results in biased classifiers that do not generalize well to test set.
However, due to the ability of our algorithms to leverage $D_T^U$, it has better generalizability.

 \etitle{Combating Overfitting.}
One issue that could reduce the performance of our algorithms in Scenarios~2 and~3 is the possibility of overfitting.
For consistency, we used the default FastText \DR size $d=300$ for all our experiments.
This could become as much as 900 for Scenarios~2 and~3 due to the transformations $\phi_S, \phi_T$, and $\phi_U$.
Given the limited amount of training data, they might not be sufficient to learn effective weights for all these dimensions.
In other words, they might overfit to the training data and do not generalize well for the test set.
While this possibility is relatively mitigated due to the nature of ER (specifically, monotonicity of precision),
it could still be a potential issue.

One approach that we found effective was to use \DRs of smaller dimension such as $d=50$ or $d=100$.
For simple datasets with limited training data, this might be preferable.
However, FastText does not provide such smaller \DRs by default.
We instead used the following approach.
We begin by taking \emph{GloVe}'s 50 dimensional word embedding dictionary.
We dump all the tuples from the datasets $D_S$ and $D_T$ into a single file such that each tuple corresponds to a sentence.
We train FastText on this dataset and require it to provide a model outputting a 50 dimension \DR.
We also instruct FastText to use GloVe's word embeddings as initial values.
This produces a \DR model that for any given word produces a 50 dimension \DR as output.

\begin{figure}[t!]
         \centering
         \mbox{
                 \subfigure[Multiple Sources]{\includegraphics[width=125pt]{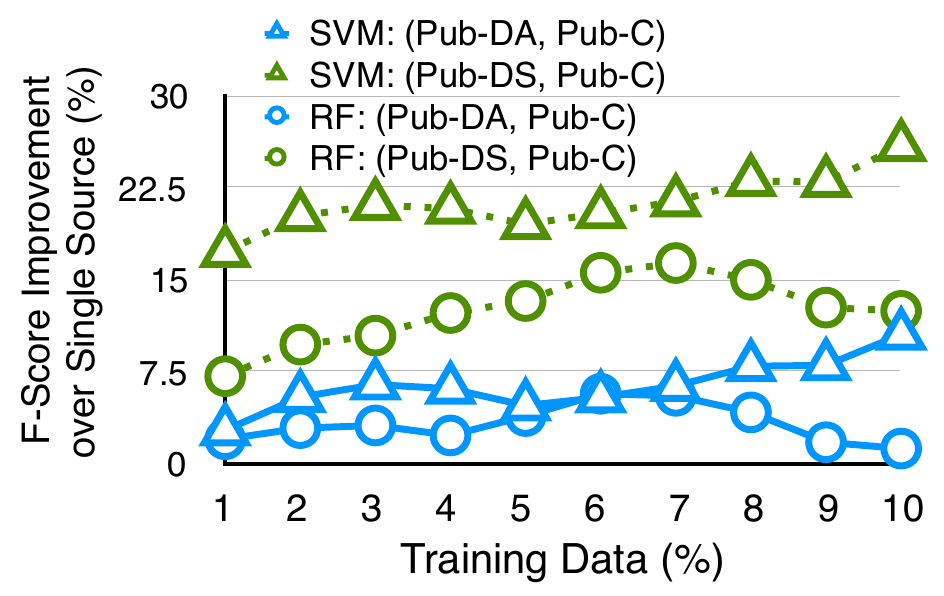}\label{subfig:multi-source}}
                 \subfigure[Cross Domain]{\includegraphics[width=125pt]{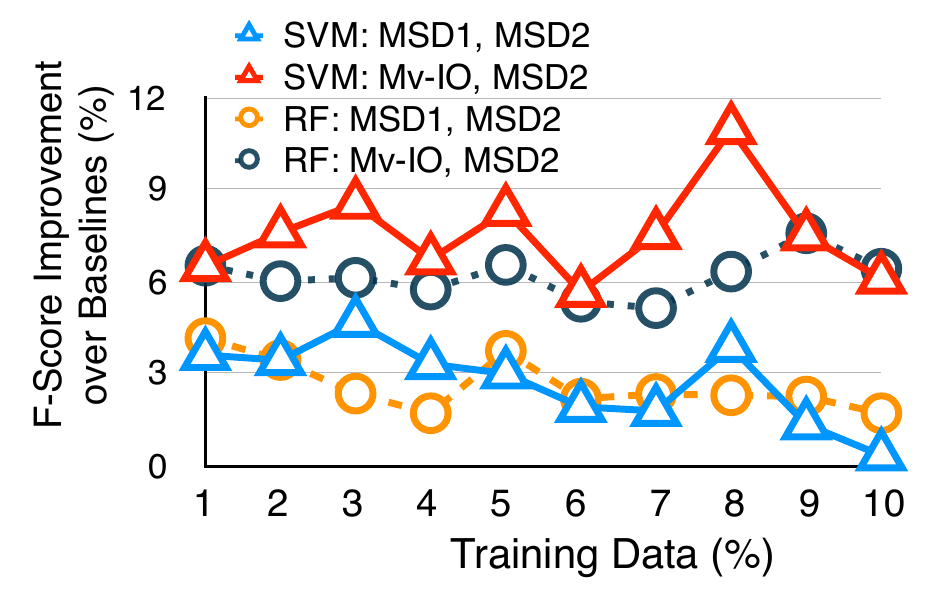}\label{subfig:crossdomain}}
         }
      \vspace*{-2ex}         
         \caption{Multiple Sources and Cross Domain}
         \label{fig:mscd}
 \end{figure}

\stitle{Exp-3: Multi Source Transfer.}
We evaluate in this experiment the potential for transferring from multiple source datasets.
We choose Pub-DA and Pub-DS as the source datasets while Pub-C is the target dataset.
We choose two representative classifiers-- SVM and RF, and evaluate three cases, namely
(Pub-DA, Pub-C), (Pub-DS, Pub-C), and (\{Pub-DA, Pub-DS\}, Pub-C).
We allocated tuples from both source datasets equally. 
For example, when we use 2\% of $D_S^L$ overall, 1\% comes from Pub-DA and 1\% comes from Pub-DS.
We measure the improvement in performance of multi source transfer (\{Pub-DA, Pub-DS\}, Pub-C) to 
single source transfers (Pub-DA, Pub-C) and (Pub-DS, Pub-C) respectively.
As expected, the improvement can be substantial (See Figure~\ref{fig:mscd}(a))
Note that Pub-DA is a better source dataset for Pub-C than Pub-DS.
So the improvement obtained in (Pub-DS, Pub-C) is as much as 25\% improvement in F-score while (Pub-DA, Pub-C) still manages to achieve 10\% improvement.
Of course, the potential for improvement is contingent on how the various source datasets are related to the target dataset. 
One can use our proposed algorithms from Section~\ref{sec:practicalIssues} for choosing one or more source datasets.

\etitle{Effectiveness of Source Selection Algorithm.}
We evaluated our source selection algorithm that ranks a given set of source datasets based on their relatedness to the target dataset.
We used a wide variety of domains and many source destination combinations.
Overall, we tested the  5 domains described earlier and 20 datasets (dubbed 784 datasets) from the Magellan data repository~\cite{magellandata}.
We compare the order provided by our algorithm 
with the empirical order where we measure the improvement provided by treating each of the datasets individually as the source.
In each case, the ordering was correct.

\stitle{Exp-4: Cross Domain Transfer.}
ML classifiers for ER take similarity vectors as input.
Once the feature space is standardized, two similarity vectors - even if they are from different datasets - can be compared.
Based on this observation, we could transfer from a dataset from a related domain whenever a dataset from the same domain is not available.
To test this, we chose two related domains - Movies and Songs.
Since Mv-IO is the larger dataset, we chose it as the source.
Recall that (MSD-1, MSD-2) is a challenging dataset due to the way the MSD dataset was partitioned adversarially.
Even in such a case, (Mv-IO, MSD-2) performed much better although it is from a different domain as shown in Figure~\ref{fig:mscd}(b). 

  \begin{figure}[t!]
  	\centering
	\includegraphics[width=160pt]{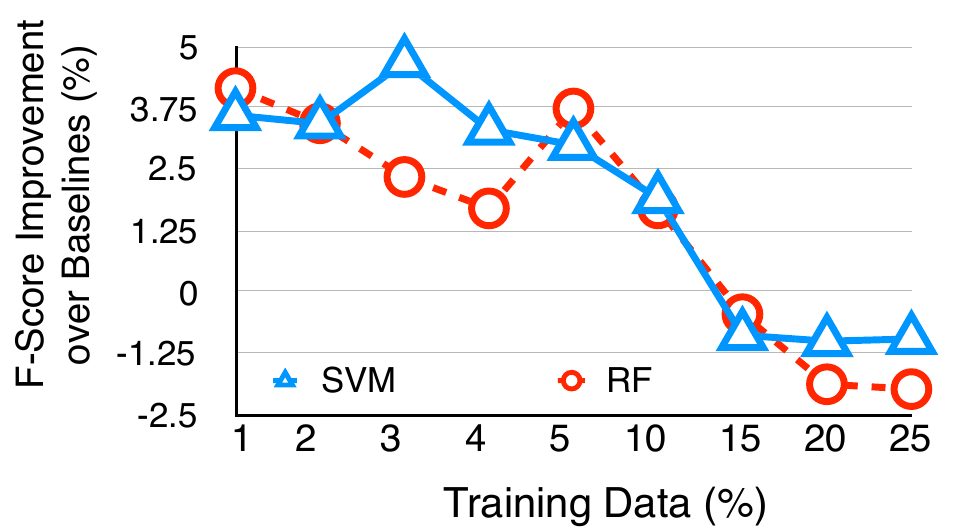}\label{subfig:negative-trasfer}
      \vspace*{-4ex}         
         \caption{Negative Transfer}
         \label{fig:negative-trasfer}
 \end{figure}
 
\stitle{Exp-5: Negative Transfer.}
It is possible that in certain cases, TL can negatively affect performance.
We constructed MSD-1 and MSD-2 adversarially to identify such a behavior.
Figure~\ref{fig:negative-trasfer} shows the results. 
We fix $D_T^L=\emptyset$ and the experiment corresponds to Scenario 1. 
We measure the improvement obtained by the instance weighting algorithm over NvT.
For small data sizes, our approach is slightly better than NvT.
As the size of the training data increases, the performance slightly improves before turning negative.

\subsection{Summary of the Experiments}

\begin{enumerate}
	\itemsep0em
	\item We have shown that TL for ER is not only feasible but also quite effective and desirable (Exp-1).
	\item Our proposed approaches provide excellent performance in all the scenarios (Exp-2).
			Furthermore, they enable ER in previously unfeasible cases such as Scenario~1
			and improve the performance in previously ineffective cases such as Scenario~3.
	\item In many cases, multi-source TL can improve the performance over a single source (Exp-3).
			We also show that our proposed heuristics (Section~\ref{subsec:sourceDatasetSelection}) can help identify the most promising sources.
	\item Our Fellegi-Sunter model based formulation 
			has an interesting side effect -- one could use the similarity vectors from a (even tangentially) related dataset
			as source $D_S$ for a given target dataset $D_T$ (Exp-4).
	\item We highlight the phenomenon of negative transfer where TL results in decreased performance (Exp-5).
		In fact, we had to partition a benchmark dataset in an adversarial manner to demonstrate negative transfer.
		We found that this occurs rarely in ER can be mitigated by techniques from Section~\ref{subsec:sourceDatasetSelection}.
\end{enumerate}

%% file: sec-relWork.tex
~\\
\section{Related Work}
\label{sec:relWork}

\stitle{(ML-based) Entity Resolution.}
ER has been widely studied by using
(a)~declarative rules~\cite{synthesizer,DBLP:conf/kdd/BilenkoM03}
(b)~ML-based models~\cite{erhard2010,deeper,konda2016magellan}, and
(c)~ experts or crowdsourcing~\cite{DBLP:journals/pvldb/WangKFF12,corleone} 
(see surveys~\cite{DBLP:journals/tkde/ElmagarmidIV07,naumann2010introduction} for more details).
 %
%
ML-based ER systems often provide the state-of-the-art performance~\cite{dong2018data,erhard2010,deeper,konda2016magellan}, by following the Fellegi-Sunter model~\cite{theoryrecordlinkage}.
Notable prior works use SVM~\cite{DBLP:conf/kdd/BilenkoM03}, 
active learning~\cite{SarawagiB02}, 
clustering~\cite{DBLP:conf/kdd/CohenR02} 
and Markov logic~\cite{SinglaD06}.
Recently, deep learning-based methods have been successfully applied for ER~\cite{deeper,anhaisigmod2018}.
These approaches leverage \DRs as the fundamental building block and propose a number of DL architectures for ER.

\stitle{ER Systems.}
A comprehensive survey of current EM systems can be found in~\cite{konda2016magellan}.
Popular tools include Magellan~\cite{konda2016magellan}, DeepER~\cite{deeper}, SERF~\cite{benjelloun2006generic}, and Dedoop~\cite{DBLP:journals/pvldb/KolbTR12}.
One of our key objectives is to adapt prior TL algorithms such that they require minimal changes to these existing ER pipelines.

\stitle{Transfer Learning} is a fundamental ML technique to improve the performance of a classifier 
on a task with limited to no labeled data by leveraging a related task (see \cite{pan2010survey,daume2006domain} for more details).
Chapter~8 of~\cite{ciml} provides a simplified version of some of the theorems that are applied in our paper.
It is possible to transfer features, training data, or ML classifiers.
We focused on the case of transferring features through distributed representations and training data.
Transferring ML classifiers is often complex and require non-trivial implementations.
Furthermore, they are also challenging to integrate into existing ER pipelines.

One prior work that uses TL for ER~\cite{negahban2012scaling}.
It focused on the specific case of multi-source ER and proposed a transfer algorithm for \emph{linear} classifiers such as SVM. However, their approach require complex modifications to the objective function and the use of composite gradient methods. 


\stitle{Distributed Representations.}
Recently, \DRs has been identified as a promising approach to encode tuples for ER~\cite{deeper,anhaisigmod2018}.
\DRs are heavily used in NLP~\cite{manning2017representations}.
Popular pre-trained word embeddings include 
word2vec~\cite{mikolov2013distributed} from Google, GloVe~\cite{pennington2014glove} from Stanford, 
fastText~\cite{bojanowski2016enriching} from Facebook,
and ELMo~\cite{peters2018deep} from AllenAI.
The definition of \DRs can be also extended from words to sentences, paragraphs and documents~\cite{le2014distributed}.
Recent theoretical works have shown that such complex methods are 
often outperformed in a TL setting by simpler compositional approaches~\cite{wieting2015towards,arora2016simple}
such as those used in our paper.

%% file: sec-conclusion.tex
\section{Conclusion}
\label{sec:conclusion}

In this paper, we have investigated the feasibility of performing ER on a dataset $D_T$ 
that has limited to no training data by reusing labeled data and features from a related dataset $D_S$. 
We proposed an effective approach to convert each tuple into a standard feature space based on distributed representations. 
We identified 4 common ER scenarios and proposed effective algorithms that can be readily integrated into existing ER pipelines.
Our extensive experiments over 12 datasets from 5 domains 
have shown that TL does provide meaningful benefits in improving the performance of an ML classifier on the target dataset.
Promising research directions to further explore include seamless integration of TL on ER with 
other research directions such as active learning, weak supervision, data/feature augmentation, and generative adversarial approaches from deep learning.